\documentclass[aps,twocolumn,pre]{revtex4-1}
\usepackage{graphicx} 
\usepackage{color} 
\usepackage{subfigure}
\usepackage{amsmath} 
\usepackage{amssymb}
\usepackage{bm}
\usepackage{exscale}
\usepackage[mathscr]{eucal}
\usepackage{upgreek}
\usepackage{color}
\usepackage{soul}

\begin{document}

\title{The frictional force on sliding drops}

\author{Joel Koplik}
\email{jkoplik@ccny.cuny.edu}
\affiliation{Benjamin Levich Institute and Department of Physics \\
City College of the City University of New York, New York, NY 10031}

\date{\today}

\begin{abstract} 
The dynamic frictional force between solid surfaces in relative motion
differs from the static force needed to initiate motion, but this
distinction is not usually thought to occur for liquid drops moving on a solid.
Recent experiments [Gao, et al., Nature Phys. {\bf 114}, 191 (2018)]
have challenged this view, and claim to observe an analog of solid-on-solid
friction for sliding drops. We use molecular dynamic simulations 
to investigate the forces that moving liquids exert on solids in several
situations.  In contrast to the indirect techniques required in laboratory
experiments, the forces involved in friction are directly accessible in
these calculations.  We find that, aside from possible inertial effects 
due to the abrupt initiation of motion and aging effects for unconfined
drops, the frictional forces are constant in time. 
\end{abstract} 

\pacs{}
\maketitle 

\section{Introduction}
\label{intro}

Liquids are often distinguished from solids by their response to shear: 
``A liquid cannot support a shear stress and flows 
irreversibly and continuously when a stress is applied'' (see, e.g.,
\cite{anderson}). 
A partially-wetting drop resting on an ideal, smooth and  planar solid surface,
held in place by surface tension forces at the contact line, would then begin 
to slide at once under gravity once the surface tilts.  Realistically however
\cite{degennes}, the drop would be held in place by surface  heterogeneities 
and only begin to move when a critical tilt angle is reached. At that point,
usually characterized in terms of the advancing and receding contact angles,
sliding begins. If instead of the liquid a second solid were placed on the
surface, once again a critical tilt angle would be needed to initiate motion
but here the usual characterization is in terms of solid and dynamic friction.  
This is a distinction between the force needed to initiate the
motion, the point where the ratio of lateral to normal force equals the 
static friction coefficient, and the lesser force required to sustain the
motion, where the ratio is the (smaller) dynamic friction coefficient.
Explanations of this phenomenon \cite{friction} involve mismatches in the
respective surface irregularities, changes in the 
degree of contact between the two solid surfaces, distortions of the solid
lattice and so on. These effects would appear to be absent in the case of a 
liquid drop, which would adjust itself to achieve complete contact with the
solid, and one would not expect a distinction between the forces (tilt angle)
needed for the initiation and maintenance of sliding motion. 

A recent experiment by Gao {\em et al}. \cite{gao} concludes that in fact 
a sliding liquid drop does exhibit distinct static and dynamic friction regimes.
In order to measure the force that a sliding drop 
exerts on a supporting solid experimentally an ingenious indirect 
technique was developed
in which a capillary pin is embedded in a liquid drop placed on a sliding
stage and the pin's deflection measured optically as the drop moves past.
Calibration of pin deflection versus applied force converts the
measured defection into the time-dependent force on the pin. 
The results indicate three
regimes: a ``static'' regime where the drop distorts but moves with the
sliding stage, a ``threshold'' regime where the drop begins to 
slip and the force on the pin rises to a peak value, followed by a
constant-force regime where the drop in held in place by the pin while
slipping over the sliding stage. This behavior was observed to be robust in 
terms of drop and solid materials and pulling speed and appears to be
general, although one may wonder about complications due to the motion of
the liquid around the pin and the consequent distortion of the drop surface.

These experimental results are quite surprising, and it would be desirable 
to have them confirmed independently by another technique.  While it is 
difficult to directly access the force exerted on an sliding liquid drop
in the laboratory, it is straightforward to do so in a molecular dynamics (MD)
simulation. In this calculation, the force between each pair of atoms is 
computed and it is a simple matter of bookkeeping to identify and isolate the 
various forces exerted on 
the drop.  In this paper we present the results of  MD simulations of
several configurations in which liquids move on solids, to test for the
presence of any analog to static solid-on-solid friction. 

What we actually measure is the force the liquid exerts on a bounding solid.  
Ostensibly, as the title of the paper suggests, we are interested in the
frictional force exerted on the liquid, but the force on the solid is both
better defined and more relevant.  
Instantaneously, the force on the liquid is just equal and opposite to the
force on the solid, by Newton's third law, but as discussed below force
fluctuations in MD are so severe that it is
necessary to average over a finite time interval to obtain a robust value 
of a force. It is straightforward to follow a region of
solid over time, since the structure is fairly rigid, and compute the force 
on that region, but liquid atoms are subject to diffusion and advection and
any liquid region will change its shape and location over time, while any
fixed region in a liquid will change its contents.  The time-averaged force
on the solid is thus well-defined whereas that on a liquid
may not be.  Furthermore, in any experiment, such as the one that motivated
this paper, it is the force on a measuring probe that is detected.

We first consider the simpler case of Couette flow initiated in a periodic 
channel by abruptly translating a bounding wall. This case has two
advantages: an analytic solution to the governing equations is available,
and there is no moving contact line issue to complicate the problem.
This simulation also allows us to identify some general features of force
measurement and the initiation of motion, which reappear in drop motion.  
Next, we study 
isolated drops sliding on a solid surface due to an applied body force.
Here we consider a partially-wetting drop which is equilibrated while resting 
on an atomistic solid surface and then subjected to a lateral body force,
in imitation of a drop placed on a tilted plane in the presence of  gravity.
We consider both a uniform atomically smooth surface as well as a
drop held in place by surface heterogeneity: an abrupt
variation in wettability or a step change in surface height. 
The force measurement results here are quite consistent with the usual
description of this flow in terms of dynamic contact angles.
Third, as a cleaner analog of the experiment of Gao {\em et al}., we study a
cylindrical drop placed in the corner where two flat surfaces intersect at
90$^\circ$, and then translate on of the surfaces so as to drive the drop
into the corner.  In this way we avoid the complicating effects of distorting 
the liquid vapor interface as the drop moves around the pin, present in the 
experiment \cite{gao}. 
Lastly we study a direct caricature of the experiment, using a
spherical cap droplet on a sliding stage held in place by a fixed pin.

In all of these simulations
we find no evidence for enhanced static solid-on-liquid forces beyond the
effects of ordinary fluid viscosity and inertia. 
In some but not all cases we observe a significant
peak in the force at early times, but we argue that this is the result of a
step change in the force applied to the liquid, which is perfectly well
accounted for by the viscous effects built into the Navier-Stokes equations,
along with some transient non-Newtonian effects.

\begin{figure}
\centering
\includegraphics[scale=0.4]{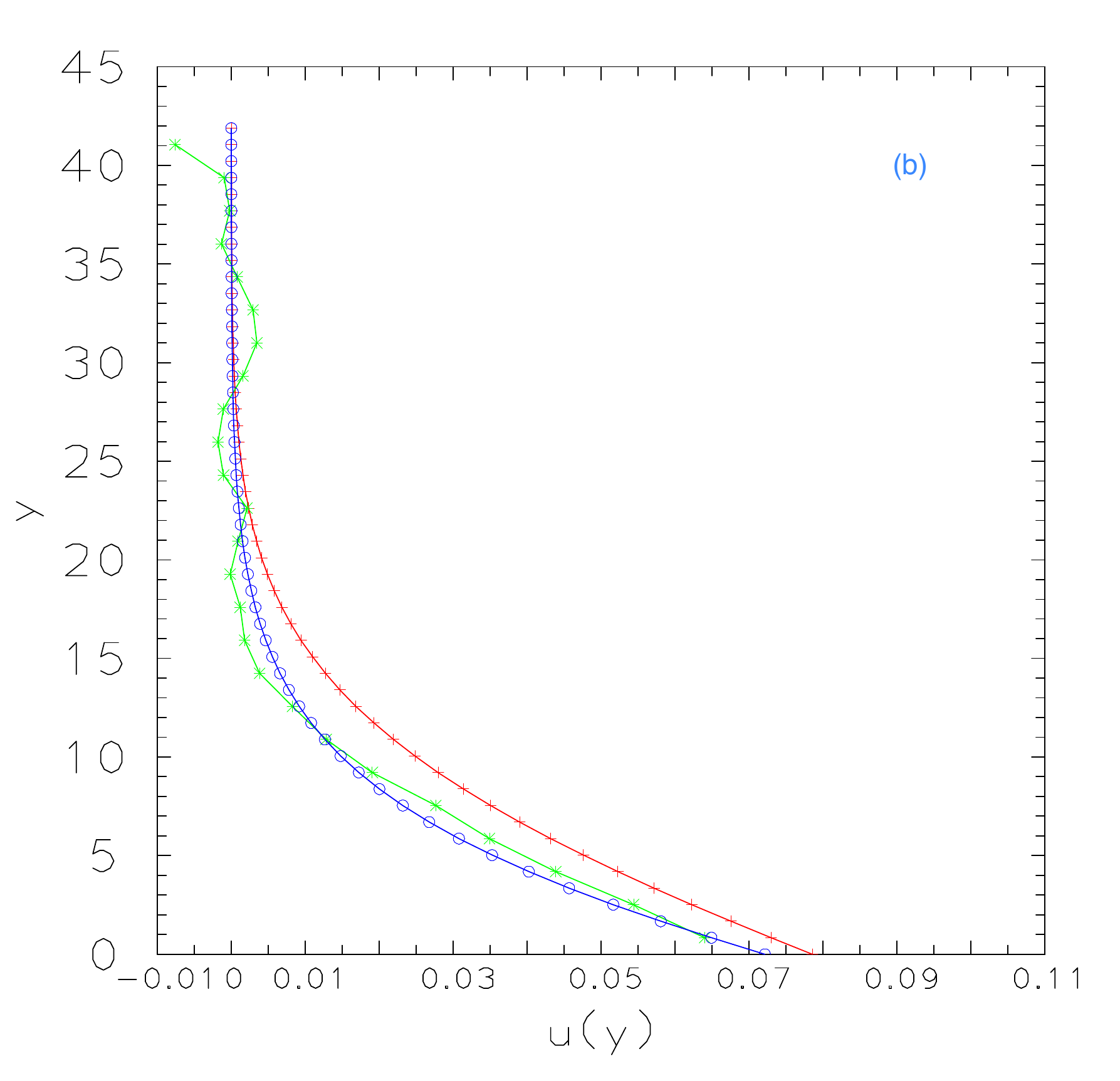}\hspace{0.1in}
\includegraphics[scale=0.4]{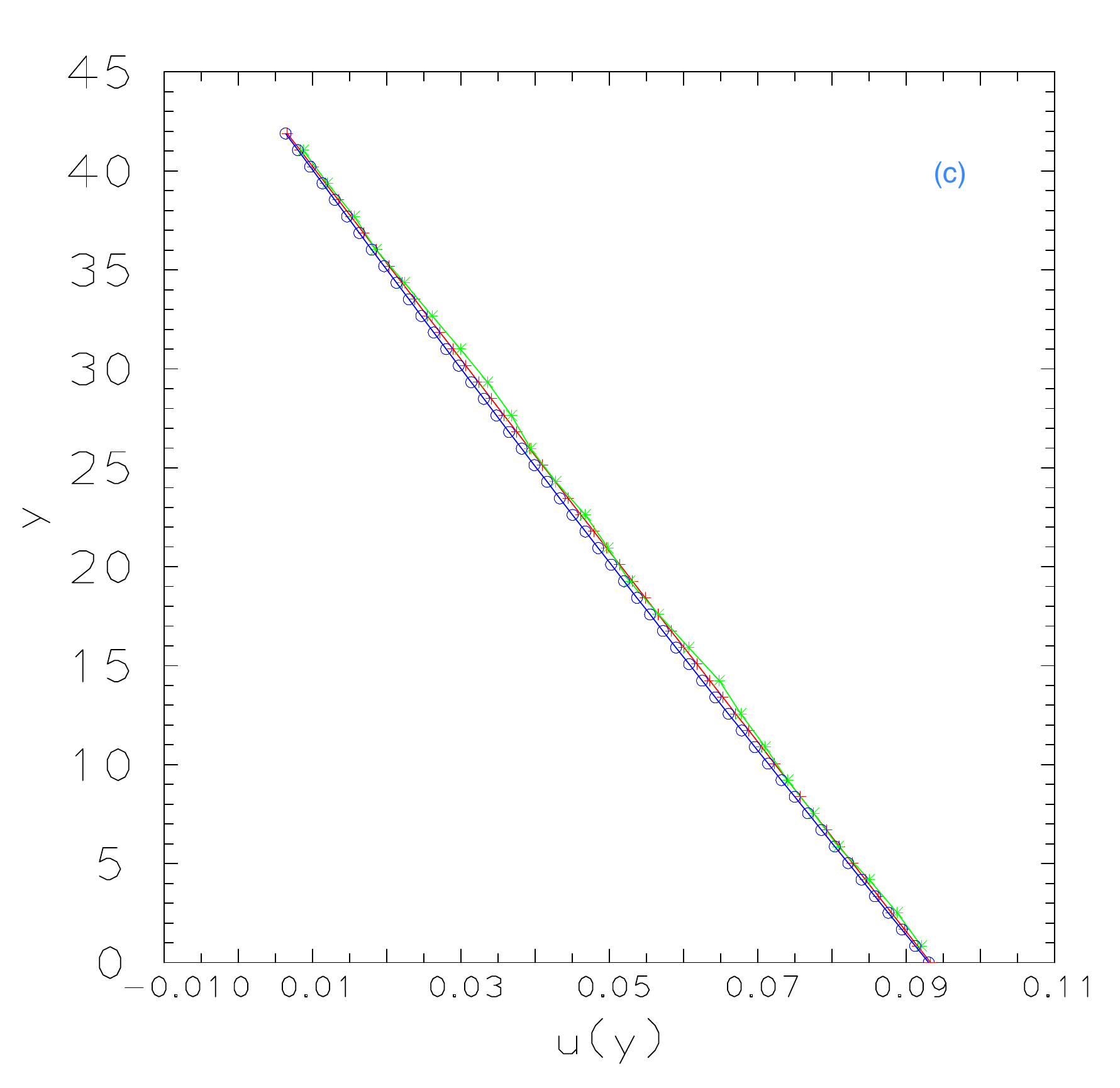}
\caption{\small{Couette flow velocity profile at 
10$\tau$ (top) and 200$\tau$ (bottom). 
The MD results (20 realization average) are in green (*), Newtonian 
Navier-Stokes in red (+) and shear-thinning model in blue ($\circ$). }} 
\label{cou}
\end{figure}

\section{Couette flow}

The simulations are based on classical MD methods \cite{allen,frenkel} and
involve a drop composed of a generic Newtonian liquid made of 
tetramer molecules with Lennard-Jones interactions, adjacent to solid surfaces
consisting of atoms tethered to lattice sites.  This computational framework
is used throughout the paper.  The interactions are 
\begin{eqnarray}
V_{\rm LJ}(r) & = & 4\,\epsilon\, \left[ \left( {r\over\sigma} \right)^{-12}
- c\, \left( {r\over\sigma} \right)^{-6}\ \right] \nonumber
\\
V_{FENE}(r) & = &-\frac{1}{2}\, k_F\, r_0^2\, \ln\left(1-{r^2\over r_0^2}\right)
\label{eq:lj}
\end{eqnarray}
with parameters $k_F=30\epsilon/\sigma^2$ and $r_0=1.5\sigma$ (after
Ref.~\cite{grest}.  In the following, if not stated explicitly it is
understood that numerical results are given in units of
$\sigma$ (length), $\epsilon$ (energy), and $\tau=m(\sigma/\epsilon)^{1/2}$ 
(time), where $m$ is the common atomic mass.
The parameter $c$ is used to adjust the strength of the liquid-solid
interaction and hence the wettability, as described below.
$V_{LJ}$, which is cut off at $r=2.5\sigma$, acts between each pair of 
atoms and $V_{FENE}$ acts between successive atoms in a four-monomer 
linear chain.  The solid atoms are bound to lattice sites using a harmonic 
potential with stiffness 100$\epsilon/\sigma^2$
and a local Nos\'e-Hoover thermostat fixes the temperature at
$T=0.8\epsilon/k_B$.
This particular liquid-vapor-solid system has been used extensively in our 
previous work \cite{jk1,jk2} and has the convenient features of 
short-range interactions, easily-variable wettability and a sharp 
liquid/vapor interface.  Furthermore, its properties have been previously
measured at this temperature:  bulk fluid density $0.857\sigma^{-3}$,
viscosity $5.18m/(\sigma \tau)$ and liquid-vapor surface tension 
$0.668\epsilon/\sigma^2$.

For Couette flow a slab of liquid is placed between two solid plates, 
15625 tetramer molecules in a cube of side $L=41.9\sigma$ with 2916 solid
atoms tethered in fcc layers above and below, and equilibrated 
for 100$\tau$.  The LJ interaction parameters are set to be $c_{ff}=1$ and 
$c_{fw}=0.75$ for fluid-fluid and fluid-wall interactions, respectively.
The bottom wall is then abruptly set into motion and translated 
steadily at velocity $U_0=0.1\sigma/\tau$. The resulting velocity profile in
the liquid at successive times is shown in the Fig.~\ref{cou} at early
(10$\tau$) and late (200$\tau$) times, and is seen to 
evolve into the expected linear form, with some velocity slip at the
walls. The wall slip occurs because the fluid-wall interaction is weak: 
the interaction parameters used
correspond to a partially wetting liquid with a sessile drop contact angle 
around 90$^\circ$. The MD results are compared to a solution of the Stokes
equation which incorporates the observed slip via a Navier boundary
condition. This continuum flow field is obtained equivalently from a 
semi-analytic solutions in the literature \cite{kg}  
or from a direct numerical integration.  The early-time agreement 
is only approximate, but this is not surprising given that we have imposed a 
shear stress instantaneously and examined the response after only about 
20 ps.  In this situation one expects some transient elastic or more
generally non-Newtonian behavior in the fluid, and specifically
a delayed response to the imposed shear.  In the literature one finds, for 
example, Couette flow calculations using FENE-P dumbbell \cite{mochimaru} 
which give a velocity profile which
lags behind the Navier-Stokes solution at early times. Furthermore, in 
Fig.~\ref{cou}b we also show a good fit to the MD velocity profile 
using a simple shear-thinning model in which the viscosity varies as
$\mu=\mu_0\cdot(1+\gamma/\gamma_0)^{-1/2}$, where $\gamma$ and $\gamma_0$ are 
the local and global strain rates.  We have also reproduced the MD profile
using a simple Maxwell model (not shown).  The point of this discussion of the
velocity profile is to argue that even for time-dependent flows, 
``experimental'' MD results for velocity are consistent with the usual 
theoretical expectations for liquids, provided one takes account of 
inertia and a possible transient non-Newtonian response to abrupt changes 
in flow conditions.  

\begin{figure}
\centering
\includegraphics[scale=0.50]{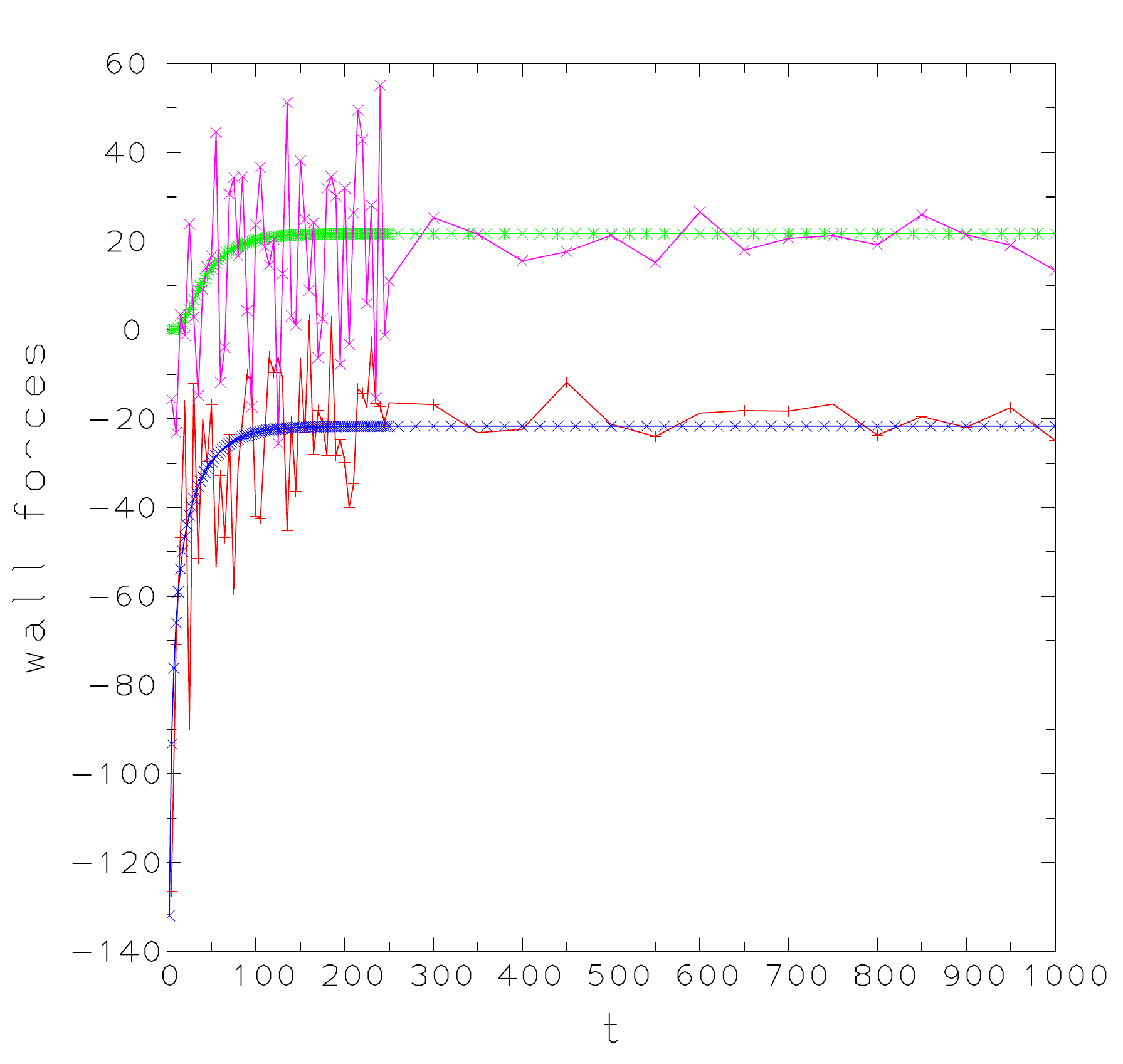}
\caption{\small{ Average wall forces in Couette flow.  
Bottom wall: MD results in red (+), Navier-Stokes in blue (x); top wall: MD
results in green (*).  Navier-Stokes in magenta (x).  }}
\label{cou-force}
\end{figure}

We now make the analogous comparison for the forces the liquid exerts on the 
solid.  In an MD calculation the force
on any liquid atom is the simple sum of pair-wise forces exerted by all 
other atoms within interaction range, so each wall force is the sum of
forces between all fluid atoms and atoms in that wall, and can be isolated 
easily.  The continuum force is just the shear stress times the
wall area, $\pm L^2 \mu \partial u/\partial y$, where the sign reflects the
direction of the normal to the interface, $u$ and $x$ are the streamwise
velocity and coordinate, and $y$ is the spanwise coordinate.  
The results for the lateral force on the two walls in Couette flow are 
displayed in Fig.~\ref{cou-force}, along with the  
corresponding continuum forces obtained from the Newtonian calculation
incorporating slip.  (The two non-Newtonian models mentioned above give
similar results.) Several features are to be noted in this figure. 

\noindent (1) The MD force is noisy, because the intermolecular force
is a rapidly varying function of interatomic spacing.  The plotted curve is 
an average over 20 statistically independent realizations, obtained from 
different initial atomic velocity values, and even then a further time 
average is needed in order to identify the trend.  The fluctuations are much
more severe at early times ($t < 200\tau$) because the data points are
spaced by 5$\tau$ and the resulting averaging interval is short, while the 
data is more stable at later times where the interval is 50$\tau$. 

\noindent (2) The continuum results are as expected.  The lower wall is
abruptly set into motion, giving a large stress at the start (nominally
infinite) which relaxes to a constant value when the velocity profile
stabilizes at linear. The liquid adjacent to the upper wall is at rest 
until the vorticity disturbance from the lower wall reaches it, so this
force is initially zero and increases to its steady value as the fluid
accelerates.  In the steady state regime the wall forces are equal and
opposite because the fluid is no longer accelerating and the forces must sum
to zero.  The force using the shear-thinning model is slightly different at
early times but still in the middle of the MD fluctuations.

\noindent (3) The MD force agrees with the continuum force, modulo the
fluctuations. The fluid begins to move immediately and the wall force does
not exhibit stick-slip behavior or any analog of solid-on-solid static 
friction.  The sharp peak in the force on the moving wall at early times
results from inertia (included in the Newtonian continuum description) plus
a bit of elasticity.    

The main issue in this paper is the possibility of a shear
stress enhancement at the onset of motion, and in this configuration the
cause is the abrupt motion of the wall. Indeed, the stress peak at $t=0$ is 
absent when the wall velocity is linearly ramped up from 0 to 0.1, in place 
of the step change illustrated in the figure. Likewise, in a similar
simulation of Poiseuille flow, even if a step pressure gradient is imposed 
the shear stress increases smoothly from zero. The distinction is
completely accounted for by the Stokes equations: the analytic solutions for
the velocity in
start-up Couette and Poiseuille flow with a no-slip boundary condition, given 
for example in \cite{pozrikidis}, directly indicate whether a stress peak
is present. These solutions involve a Fourier series which 
converges for the shear stress at the bottom wall at $t=0$ for Poiseuille
flow but diverges in the Couette case. 

\section{Sessile drops}

\begin{figure}
\centering
\includegraphics[scale=0.25]{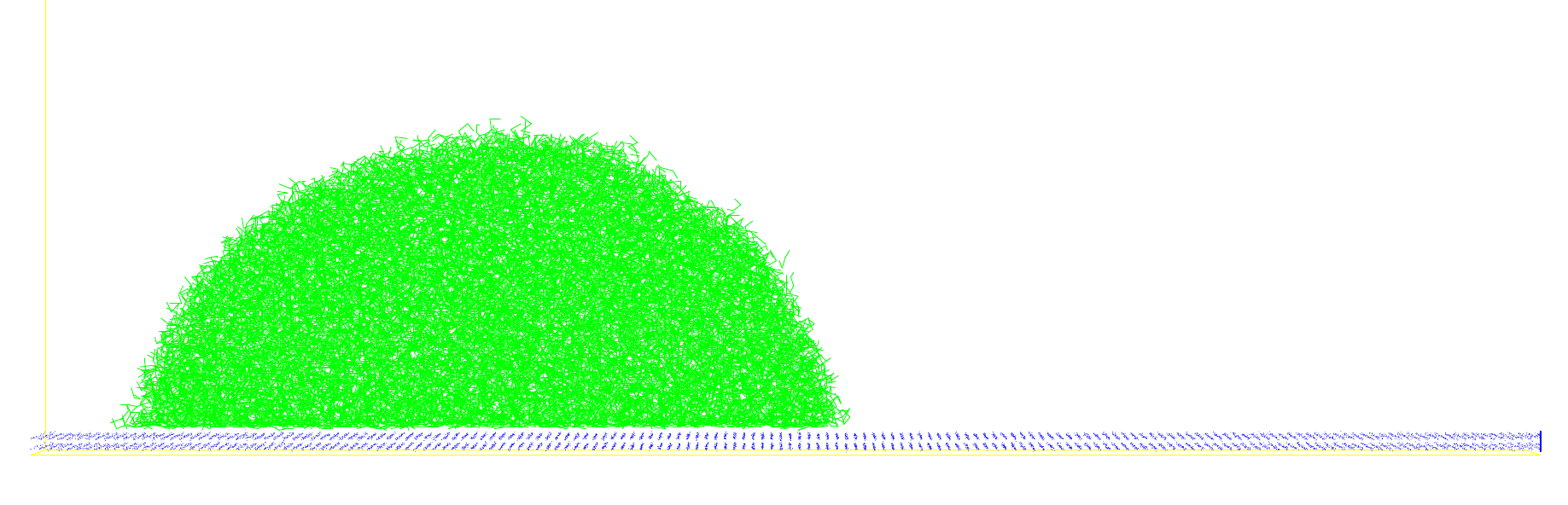}\hspace{0.1in}
\includegraphics[scale=0.25]{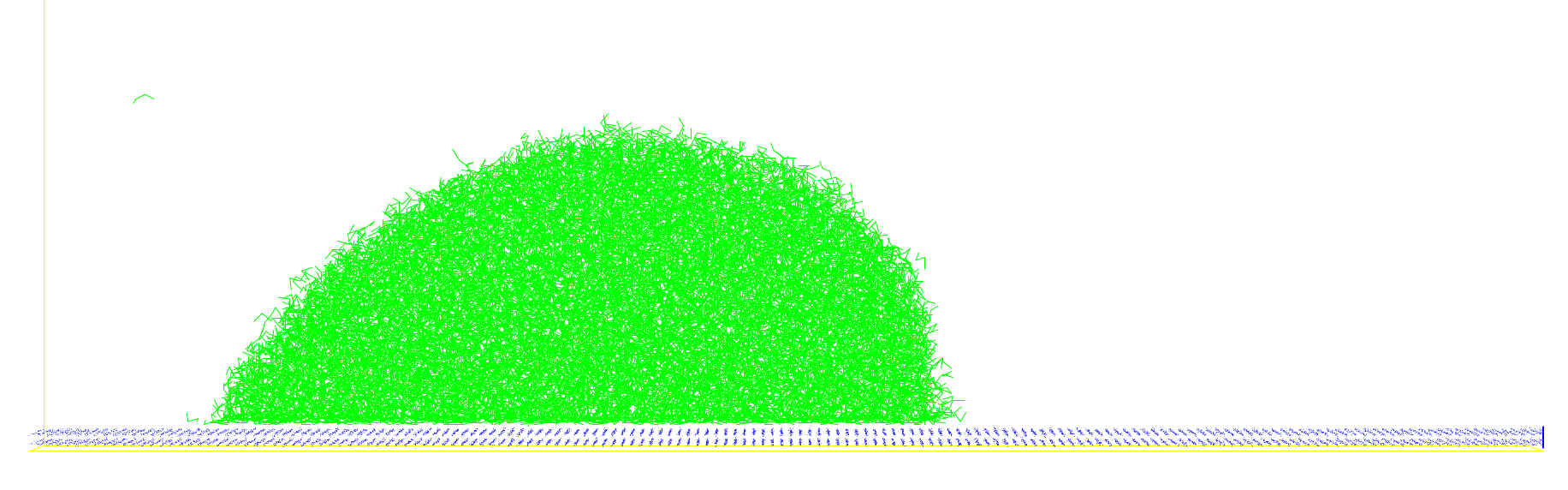}\hspace{0.1in}
\includegraphics[scale=0.25]{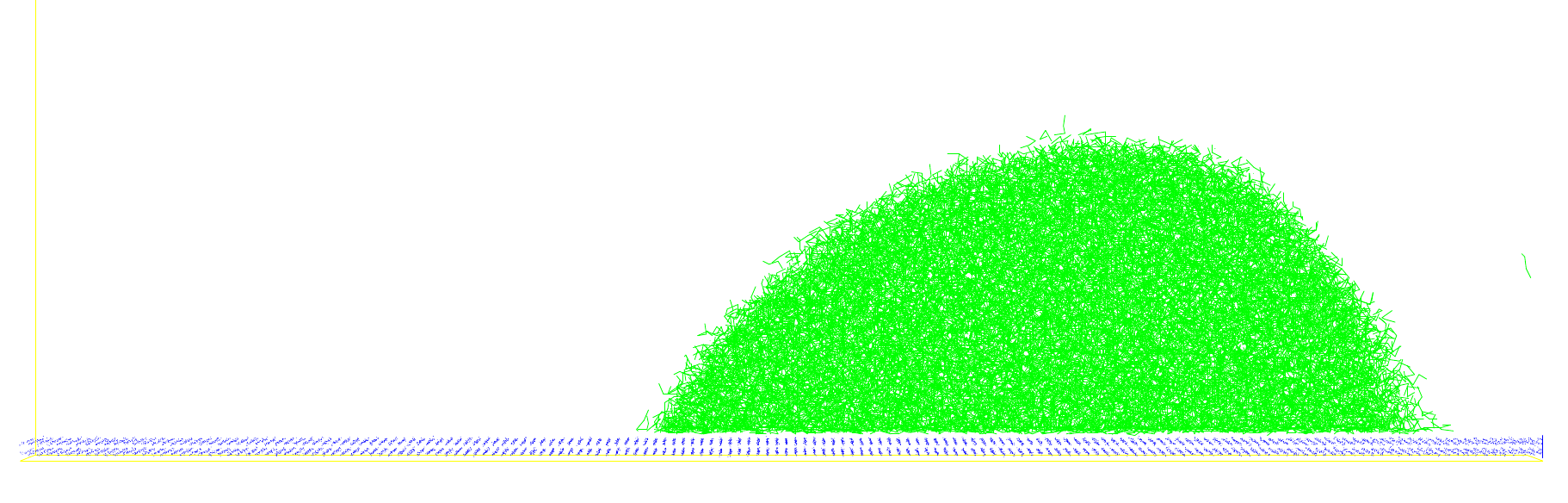}
\caption{\small{ Sessile drop driven across an atomically-smooth 
homogeneous substrate at times 0, 500$\tau$ and $2500\tau$ (top to bottom). }}
\label{smooth-snap}
\end{figure}

The simplest fluid mechanical analog of solid-solid friction involves a 
drop sliding on a solid surface.  Although these is an extensive literature  
on this problem \cite{spread1,spread2} relatively little attention has been 
given to the
forces involved, and we use MD simulations to disentangle them.  We begin
with a partially-wetting liquid on a uniform and atomically-smooth substrate,
in the form of cylindrical cap drop shown in Fig.~\ref{smooth-snap}.  
The advantage
of this shape compared to a spherical cap is that the system is statistically
homogeneous in the direction ($z$) normal to the plane of the figure, and the
results can be averaged over $z$.  The wall-fluid interaction strength is
chosen as $c_{ff}=0.85$ which gives an initial contact angle of 70$^\circ$.
After equilibration a constant force 
$f=0.001 m\sigma/\tau^2$ is applied to each atom and the drop
translates, as shown in the subsequent frames of the figure. 
The Reynolds number, 
$Re\equiv \rho \dot{X} H_0/\mu$ where $H_0$ is the equilibrium
drop height and $\dot{X}$ is the center of mass velocity (see below), 
is 0.0748 and the Capillary number $Ca\equiv\mu\dot{X}/\gamma$
is 0.149, with slightly different values in the other cases. 
As the drop moves its shape changes and exhibits distinct  
advancing and receding dynamic contact angles which, as seen in 
Fig.~\ref{smooth-snap}, fluctuate during the motion.  A constant
external force is formally equivalent to tilting the substrate in the
presence of gravity, although realistically the values of $f$ used here are
much too large for this interpretation.
The behavior is similar over a range of forcing values near this one,
$5\times 10^{-5}$ to 0.005, but for
significantly lower values the force on the drop is unmeasurable because the
signal is swamped by the
fluctuations and the drop motion itself becomes intermittent. As for
higher forcing values, 
at $f=0.01$ the center of mass still moves with constant velocity although
the drop becomes increasingly elongated and at still
higher values the drop tends to fly off the substrate. 

\begin{figure}
\centering
\includegraphics[scale=0.40]{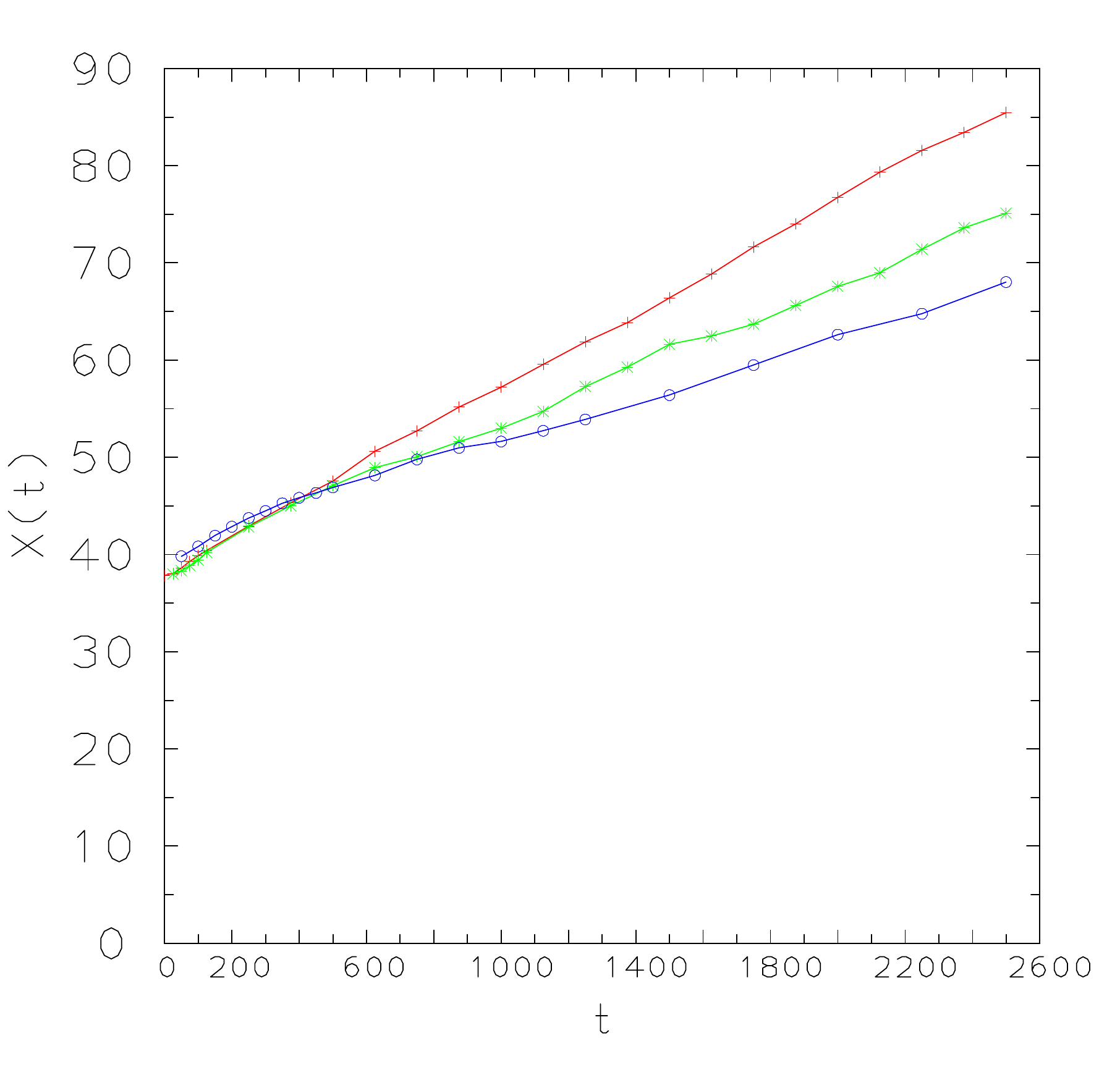}\hspace{0.1in}
\includegraphics[scale=0.40]{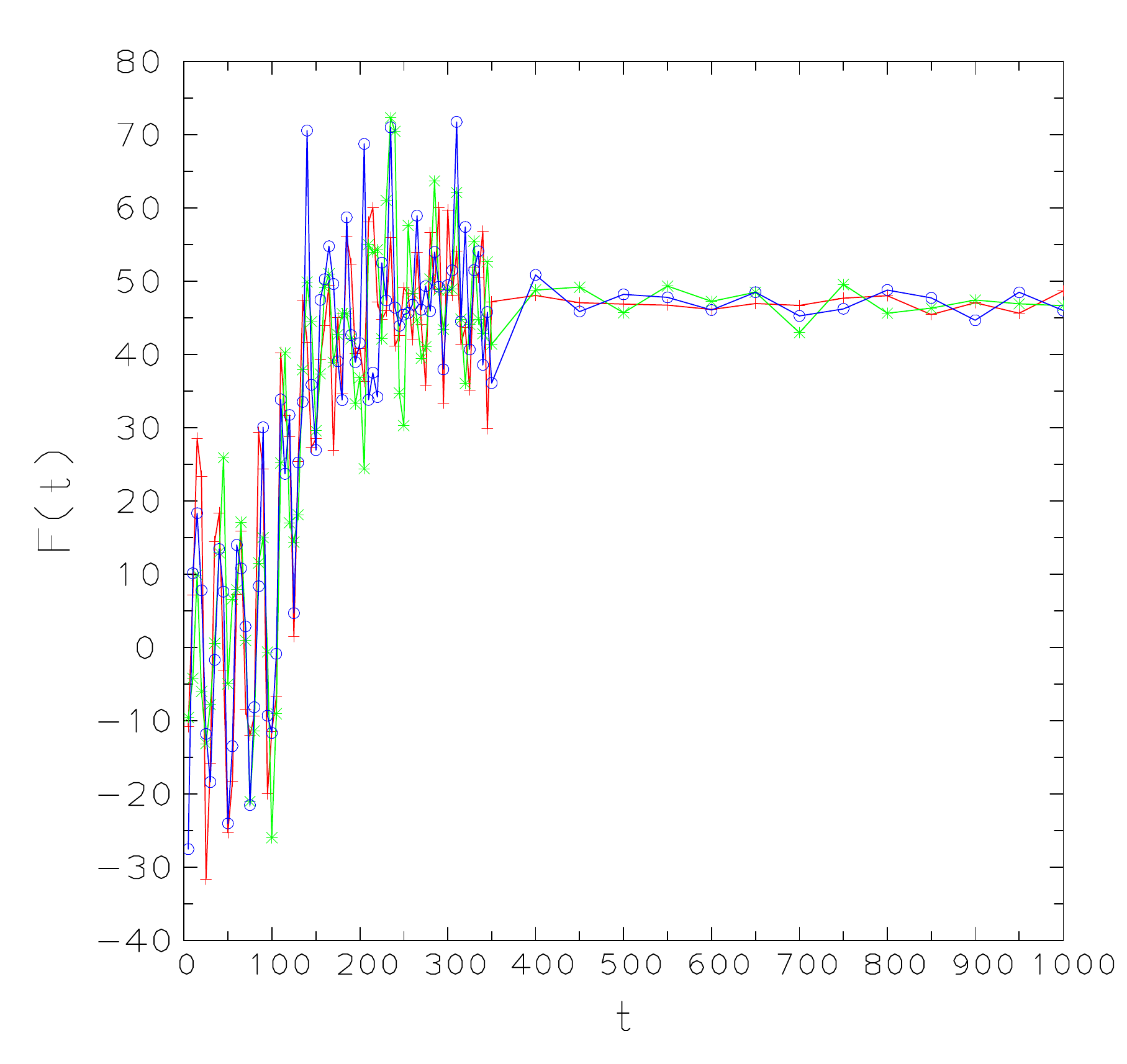}
\caption{\small{ Center of mass (top) and lateral force (bottom) vs. time 
for a sessile drop forced across the various substrates at $f=0.001$, 
averaged over 20 realizations. Smooth: 
red (+), chemical step: green (*), physical step: blue ($\circ$). }}
\label{interior}
\end{figure}

\begin{figure}
\centering
\includegraphics[scale=0.25]{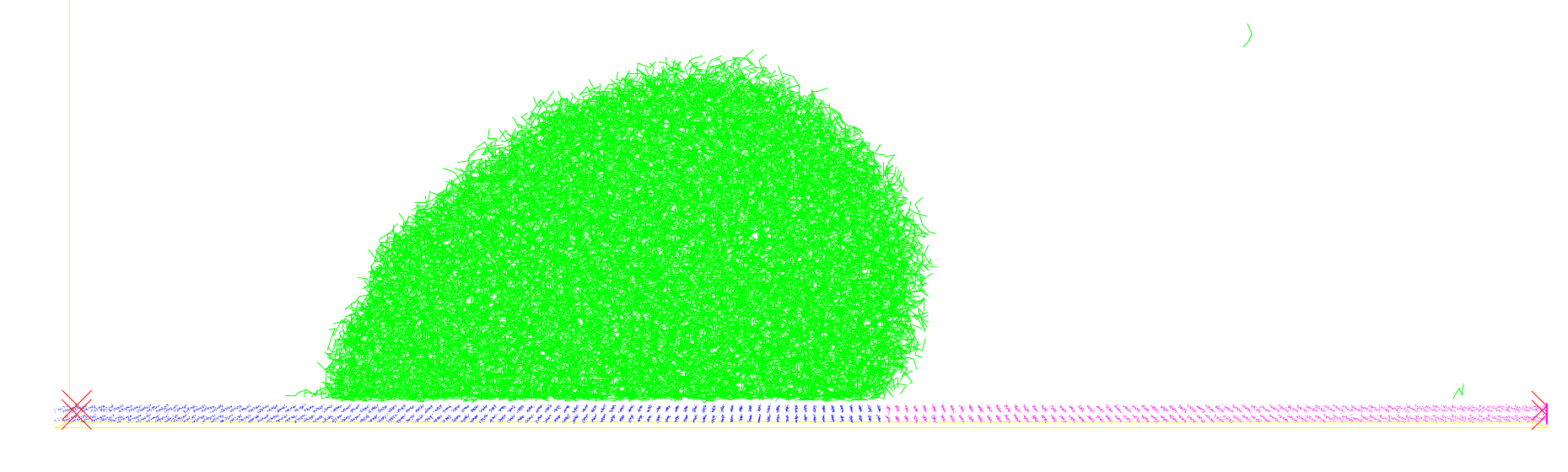}\hspace{0.1in}
\includegraphics[scale=0.25]{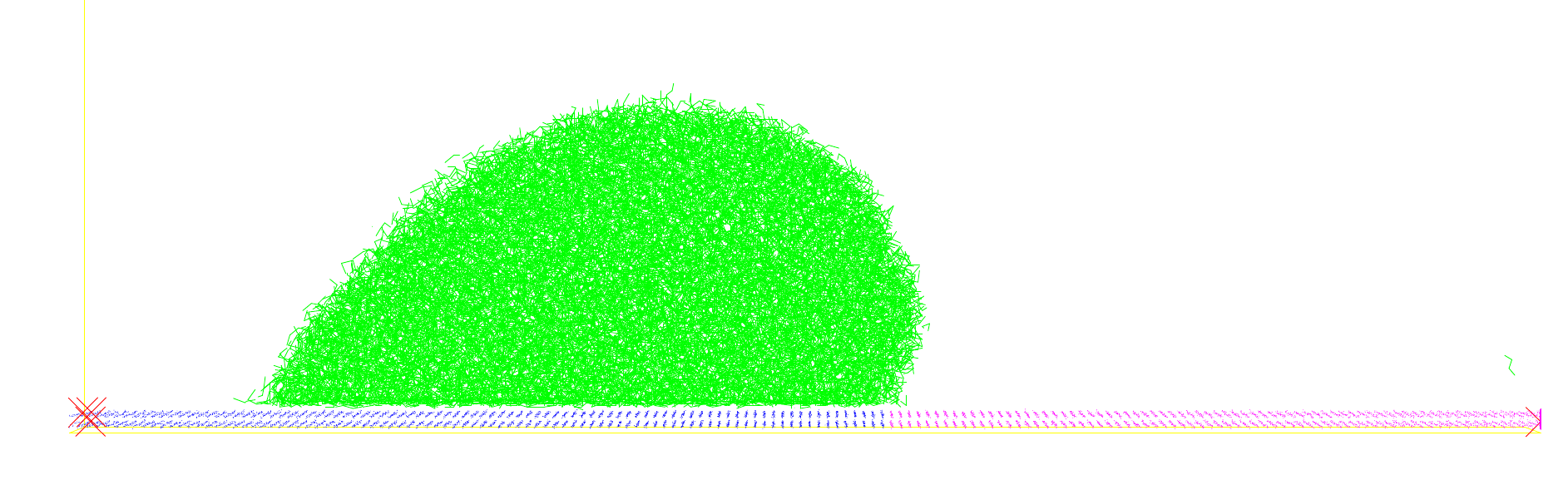}\hspace{0.1in}
\includegraphics[scale=0.25]{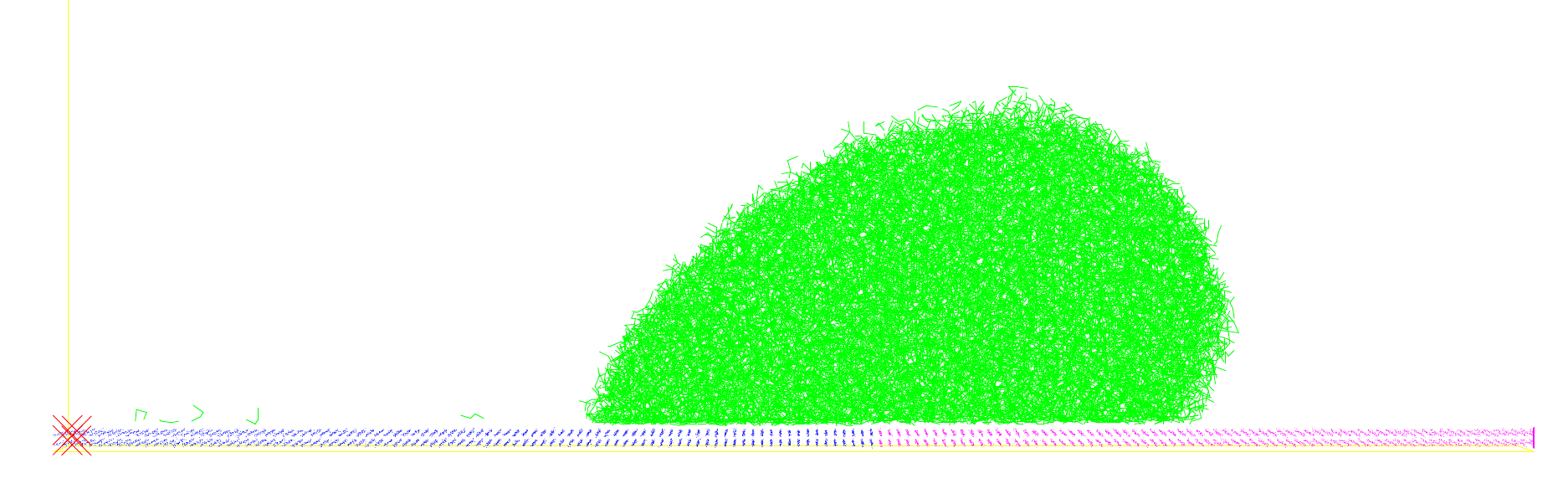}
\caption{\small{ Sessile drop driven across an 
substrate with a chemical step. Pinned at $f=0.0005$ (top) and in motion 
for $f=0.001$ at times 500$\tau$ (center) and $2500\tau$ (bottom). }}
\label{chem-snap}
\end{figure}

Most realistic solid surfaces are heterogeneous and irregular, and we
can address these complications by forcing the drop across a modified 
substrate involving either a chemical or a physical heterogeneity.
The drop is equilibrated as before but 
to the right of the drop there is either a chemical step where 
the wettability coefficient $c_{fw}$ drops to 0.65 or a physical step 
where the height
of the solid rises by one fcc unit cell, a distance of 1.17$\sigma$ here. 
When a lateral force is applied the drop tilts in the direction of the force,
developing distinct advancing and receding angles, $\theta_{A,R}$,
but moves only if the 
force is large enough. We illustrate this behavior in Fig.~\ref{chem-snap} 
for the chemical step case, where the value of the threshold force needed 
to move the drop over the step can be obtained by a the following force balance
\cite{degennes}.

In equilibrium on a homogeneous surface a drop is held in place by 
solid/liquid, solid/vapor and liquid/vapor surface tension forces at the 
contact line, with a lateral force balance given by Young's equation
$\gamma_{SV}-\gamma_{SL}=\gamma\cos{\theta_0}$. For non-ideal surfaces
there is a  range of possible equilibrium contact angles. If a weak
force $f$ is applied to each atom, a drop tilts in the
direction of the force but can remain at rest if the
applied force is balanced by the unbalanced surface tension forces per length 
$\gamma(\cos{\theta_R}-\cos{\theta_0})$ at the receding contact line and 
similarly at the advancing line.  Explicitly, the force balance is
$fN=\gamma W(\cos{\theta_R}-\cos{\theta_A})$, where $N=46800$ is the
number of atoms in the drop and $W$ length of the contact line (the width of 
the drop in this situation, $51.3\sigma$).  For the chemical step, we 
observe that the drop 
appears stationary at force $0.0005$, where it is pinned at the edge of the
step, but moves steadily at 0.0006 and higher values. 
Using the drop density profile at 0.0005 and defining the liquid/vapor 
interface to be
the contour where the density is half the bulk value, we estimate 
$\theta_R\approx 67^\circ$ and $\theta_A\approx 111^\circ$, giving $f=0.00055$ 
from the force balance
equation, in agreement with the simulation. A similar transition and
corresponding force balance is found in the case of a physical step.
If there are no angles $\theta_{A,R}$ in the equilibrium range satisfying the 
force balance the drop will move, and in the present simulations moving drops
are expected to have have constant velocity, characteristic of linear friction 
in Stokes flow. 

We observe that, despite the shape fluctuations seen in the figures,
the motion is steady: for forcing $f=0.001$ 
Fig.~\ref{interior} shows the drop center of mass $X(t)$ moving with nearly 
constant velocity from the start for both smooth and heterogeneous surfaces. 
The total lateral force $F(t)$ the liquid exerts on the wall at $f=0.001$ is 
given as a function of time in Fig.~\ref{interior}.
in all cases the wall force oscillates about zero before the external 
force is applied at time 100$\tau$ and
afterwards ramps monotonically up to to a plateau. 
Only the transient behavior immediately after application of the external 
force shows a variation with the structure of the substrate.  The
distinction between the size of the fluctuations at early and late times is
the same as in the previous Couette simulations: short and long
averaging intervals, respectively.
The average force in the plateau fluctuates about the value of the net 
force applied to the liquid, $Nf=46.8$, 
as expected because the drop is not 
accelerating. It is possible to decompose the wall force into
capillary forces at the contact line plus frictional drag on the drop, but
unfortunately the dynamic contact angles could not be measured with any
accuracy, and we have not pursued this.

\begin{figure}
\centering
\includegraphics[scale=0.40]{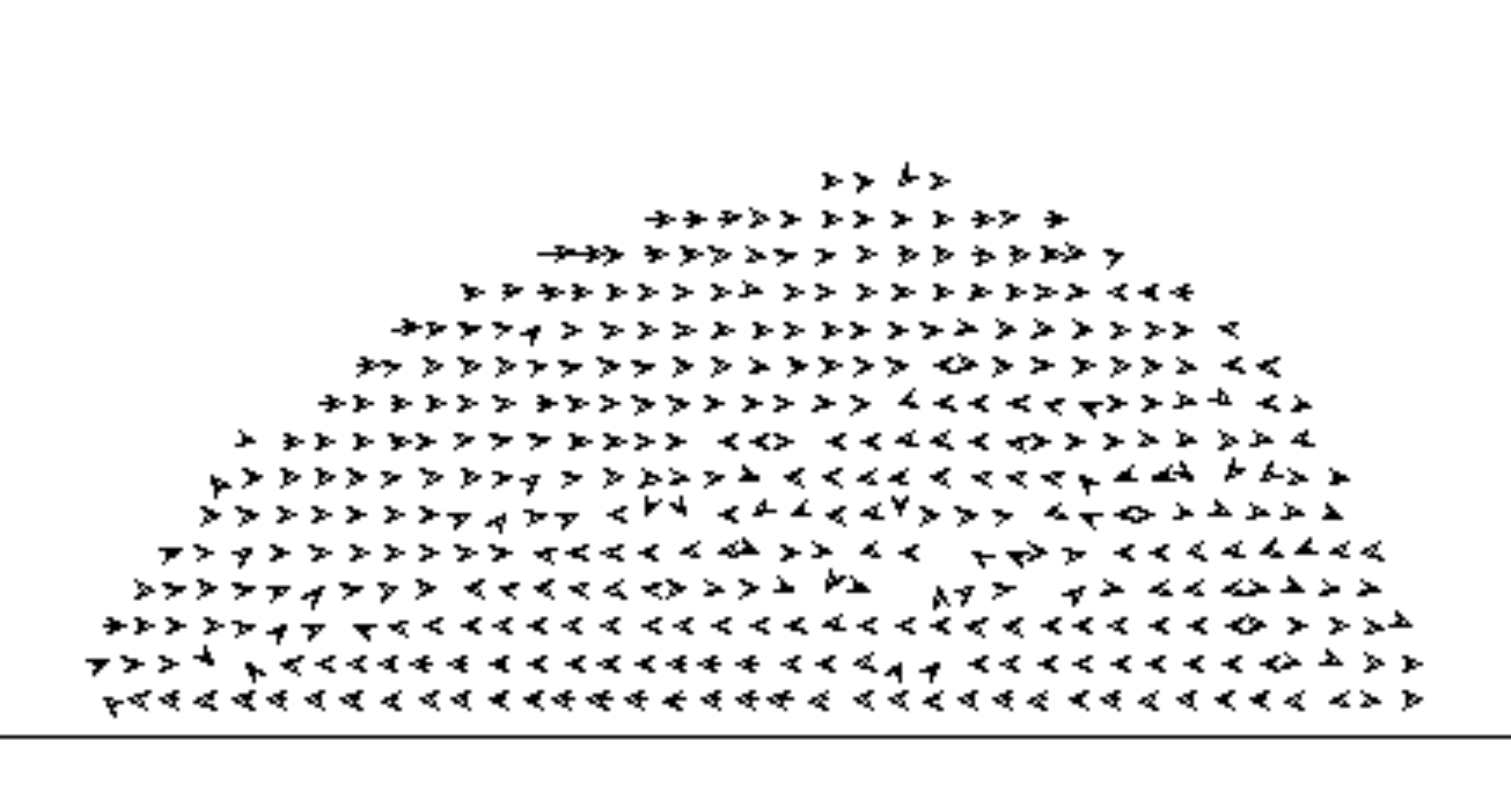}\hspace{0.1in}
\includegraphics[scale=0.40]{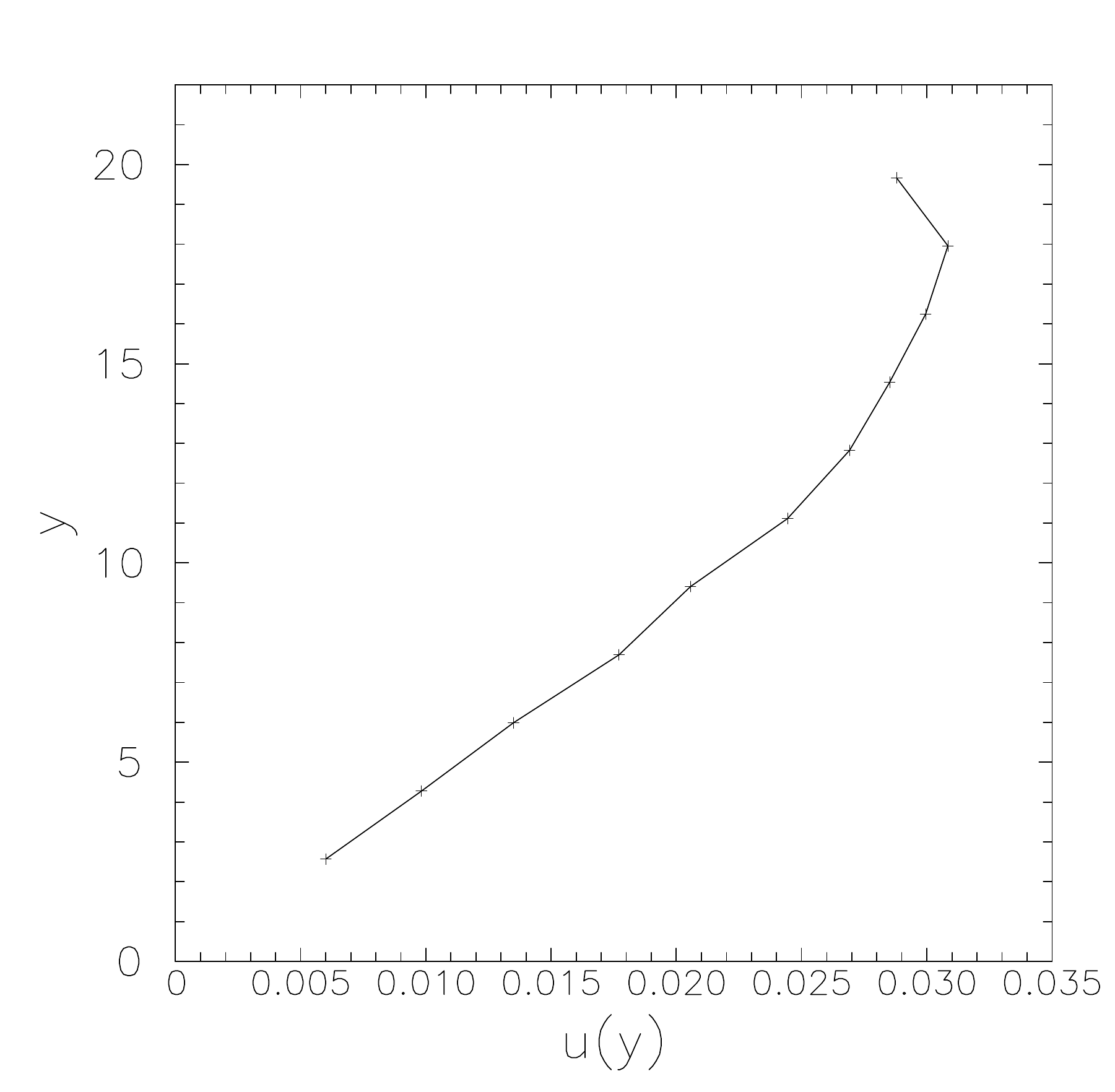}
\caption{\small{ Flow inside the drop on a homogeneous substrate. 
Two-dimensional velocity field in the moving center of mass frame (top) and
average lateral velocity profile
in the center of the drop in the substrate frame (bottom). }}
\label{roll}
\end{figure}

The motion of the liquid inside the drop is a combination of rolling and
slipping, as indicated in Fig.~\ref{roll} for the smooth wall. 
The two-dimensional flow field
shown is an average over 50$\tau$ and evaluated in a reference frame moving
with the drop center of mass; the resolution is poor since the velocities
involved are $O(10^{-2})$ times the random thermal velocity and the
center of mass velocity is not exactly constant during the averaging interval, 
but rotation
about the middle of the drop is evident.  Ensemble averaging tends to wash
out the result for this two-dimensional field, but is more effective for the 
velocity profile, which is instead
evaluated in the reference frame of the solid wall and is also an average
over the middle of the drop. The result is a roughly linear profile as would 
correspond to rotation about the wall. (The uppermost points correspond to
the liquid/vapor interfacial region where the density is falling off.)
The presence of slip at
the wall requires a small discussion. The region $0\le y\le 1.71\sigma$ is
occupied by solid (at rest), and the lowest liquid point (at $y=2.565$)
is in the center of a finite-sized sampling bin in the liquid. The precise
definition of the ``solid wall'' is always ambiguous at atomic resolution,
but would certainly be somewhere in the region $1.71\sigma\le y\le 2.5\sigma$,
where $u$ is non-zero. Aside from this ambiguity, as in Couette flow the 
presence of slip varies with the strength of the fluid-wall interaction  
but the rolling motion is always present.

\begin{figure}
\centering
\includegraphics[scale=0.40]{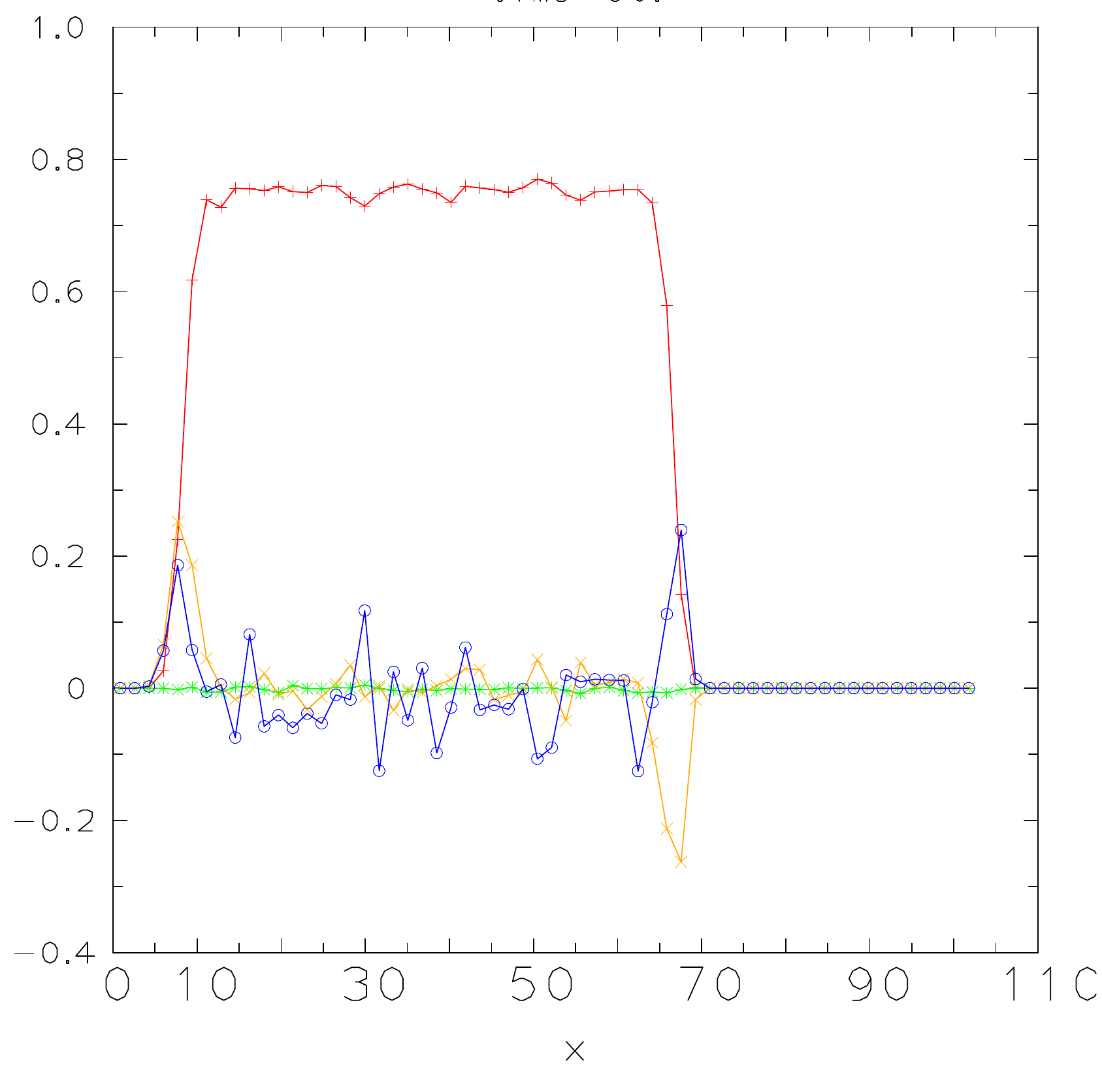}\hspace{0.1in}
\includegraphics[scale=0.40]{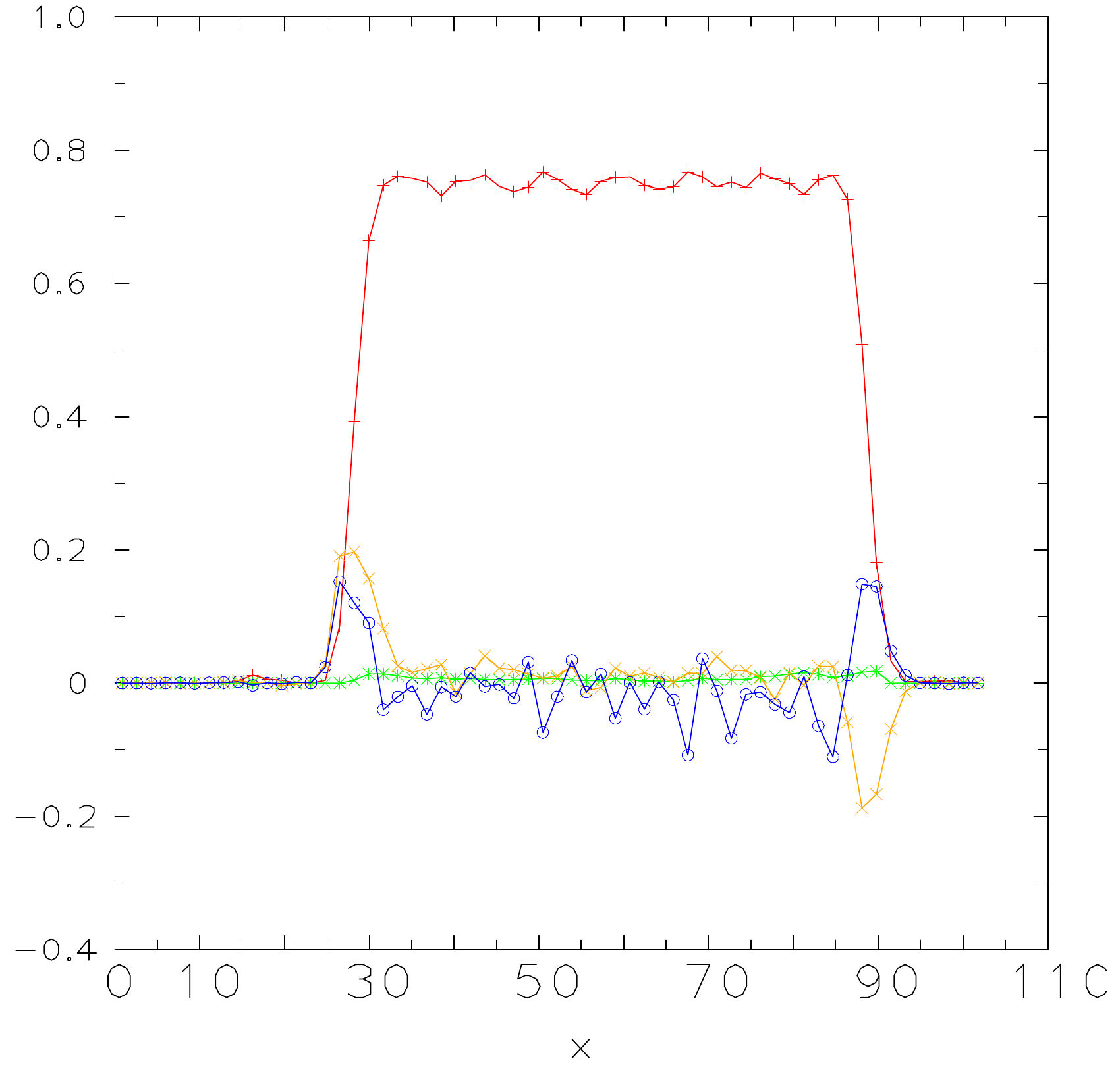}
\caption{\small{ Local fields for a drop on a homogeneous substrate,
at rest (top) and in motion (bottom) after 250$\tau$, in a single
realization.  Density: red (+),
$x$-velocity: green (*), $x$-force: orange (x) and $y$-force: blue
($\circ$). }}
\label{bottom}
\end{figure}
 
The force exerted on the drop by the walls, can be examined {\em locally}
by computing the force as a function of the coordinate along the base of
the drop.  We use a
sequence of slabs of wall, $i*dx<x<(i+1)*dx$ and $0<y<L$ with
$dx=1.71\sigma$, and compute the force per area on each slab.  (Note that
this quantity is {\em almost} the $x$-$y$ component of the fluid's shear
stress tensor at the wall. The qualification is because, as in the slip 
discussion above, MD fields are always averages over a finite-sized sampling 
bin with the result assigned to its center, which is displaced from the
wall by half the bin size. A further extrapolation would be required to 
determine the force at the wall.) 
In Fig.~\ref{bottom} we show the horizontal ($x$) and vertical
($y$) components of the force on the drop, along with the local density and
lateral velocity, both in equilibrium and while translating. The forces peak
at the drop's two contact lines and are constant on average in the interior.
In the $x$-direction, the signs of the two peaks correspond to the fact
that the liquid/vapor surface tension acts to contract the drop into to a 
circular cylinder but the attraction to the wall draws the edges of the drop
outwards, corresponding to an inward force {\em on} the wall. 
The wall force in the interior of the drop is (statistically) constant, zero
when the drop is in equilibrium and positive during drop translation due to
the liquid pulling to the right. In the $y$-direction, the force peaks at
the contact lines, again correspond to surface tension trying to contract the
drop by pulling upwards on the wall, while in the interior the force
is a constant, corresponding to the Laplace pressure inside the drop.

A similar but more detailed discussion of the variation of these local 
forces with contact
angle and a comparison to Young's equation is given by Fernandez-Toledano, 
{\em et al}., \cite{ft} for a different configuration involving a liquid 
bridge in equilibrium between two
solid plane walls.  In that case the drop is confined by the walls but here
is it is only attracted to one wall, so the force analysis differs in detail. 
The principal point we wish to make here is that when the drop moves
the numerical values change but there is no qualitative
difference in local forces on the wall.  The density is constant inside
the drop, and its profile simply shifts in $x$ as the drop moves, while the
liquid velocity at the wall is constant, zero in equilibrium and
approximately equal to the center of mass velocity during the motion.

\begin{figure}
\centering
\includegraphics[scale=0.25]{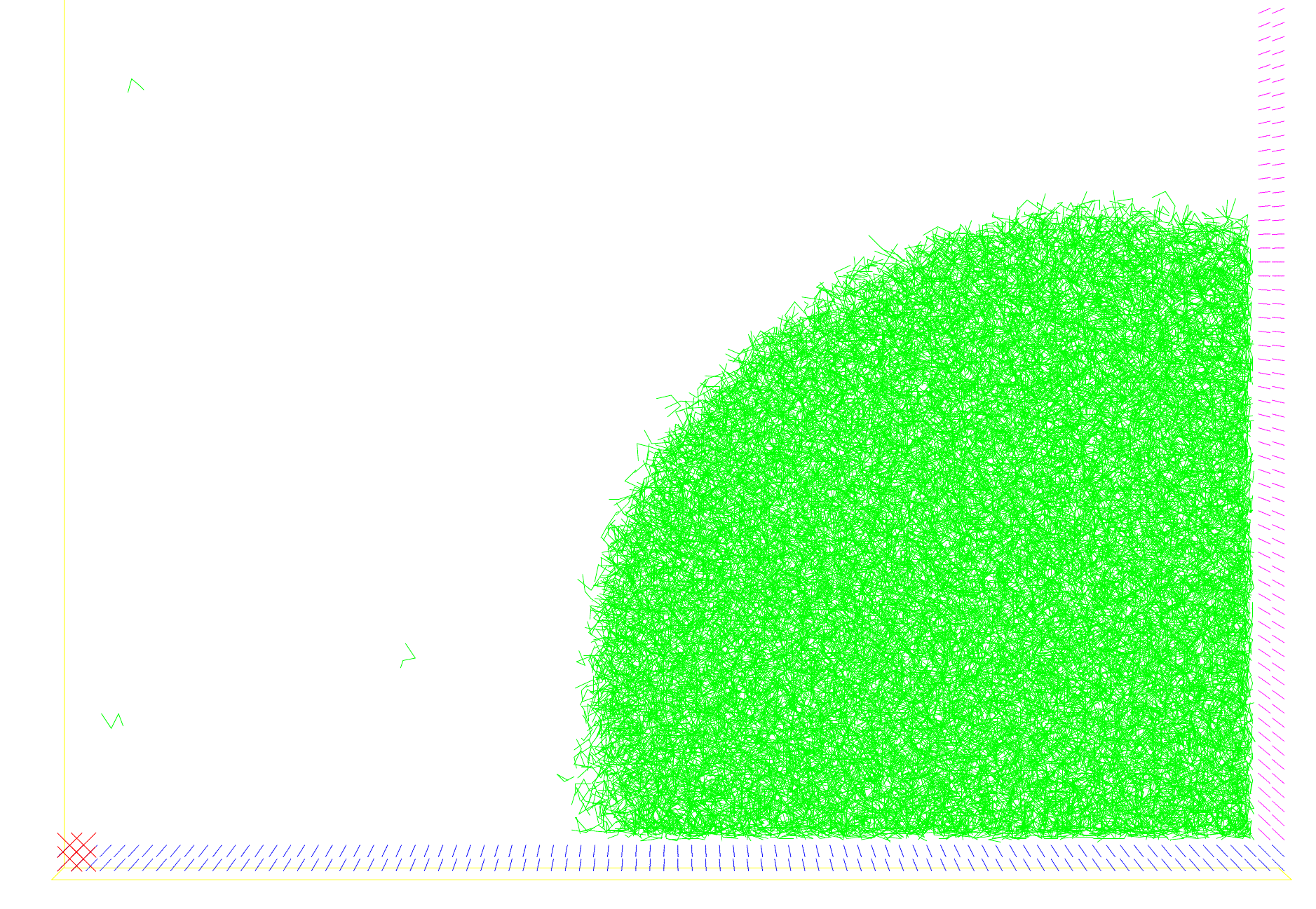}\hspace{0.1in}
\includegraphics[scale=0.25]{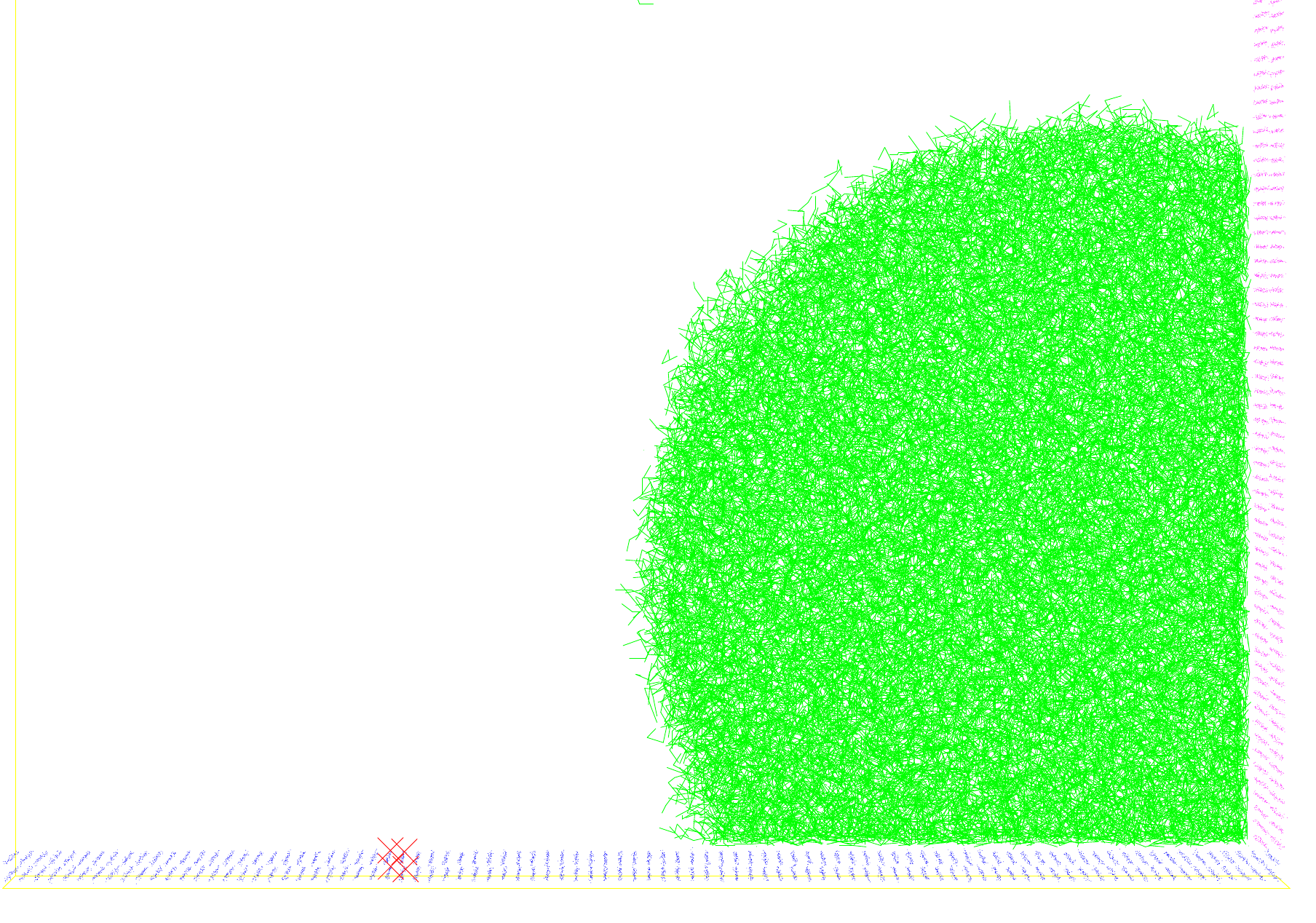}\hspace{0.1in}
\includegraphics[scale=0.25]{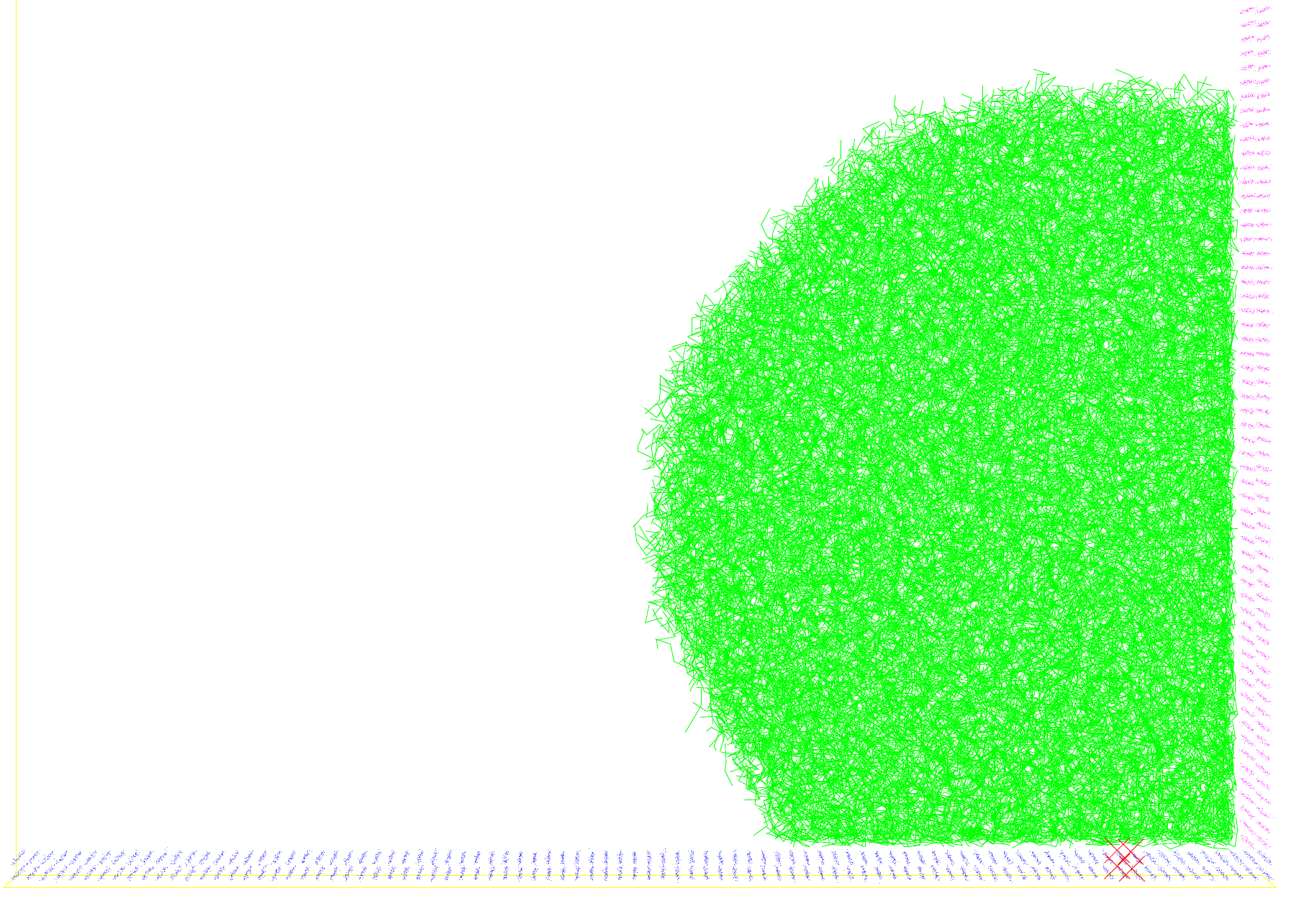}
\caption{\small{ Drop driven into a corner by motion of the bottom
wall to the right, at times 0, 200$\tau$ and $600\tau$ (top to bottom). 
The translation of
the wall is shown by the red dot which indicates the location of an
atom fixed in the wall,  }}
\label{snap-corner}
\end{figure}

\section{Cornered drops}

The experiments which motivated this study involved spherical drops on a 
sliding stage
held in place by a measurement pin, and includes features of both of the
previous simulations.  The liquid is forced into motion abruptly as in
Couette flow, with a localized impulsive stress at a solid boundary, and 
then ``slides'' relative to the
solid with (as we shall see) a combination of slip and rolling.  Again
we take advantage of MD's direct access to the force and use a simpler
configuration with no variation in one direction to improve the statistics.

We place a drop in the corner formed by two distinct solid walls meeting 
at a right angle, as shown in Fig.~\ref{snap-corner}, and choose the
(partial) wetting coefficient $c_{ff}=0.75$.  The simulation box is 
a cube of side 68.4$\sigma$ containing 16,000 tetramer molecules, which
initially fill an approximate circular quadrant with a contact 
angle of 93$^\circ$.  The vertical
wall is fixed in place while the horizontal wall is, after equilibration, 
translated towards the fixed wall at a constant velocity. Due to
periodicity, the sliding wall can translate indefinitely. The gap where the
two walls meet is the same size as the internal wall lattice spacing, and there 
is no interaction between the respective wall atoms, and therefore no
fluid leakage or distortion of the corner or any solid frictional heat 
generated there. These simplifying properties would be difficult to realize
in a laboratory experiment, although the differences involve only the 
corner region which is not the focus of this study in any case. 

\begin{figure}
\centering
\includegraphics[scale=0.45]{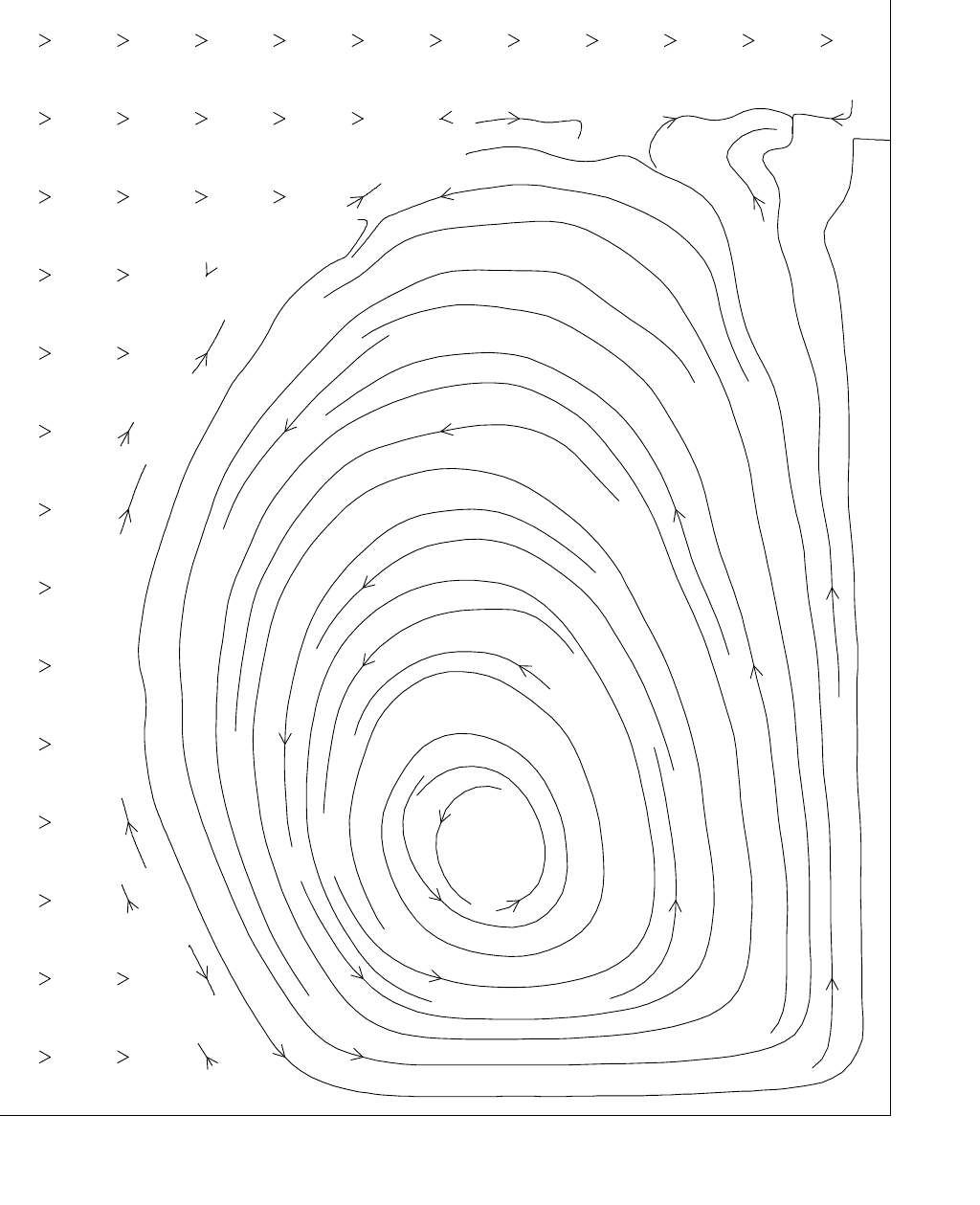}\hspace{0.1in}
\includegraphics[scale=0.40]{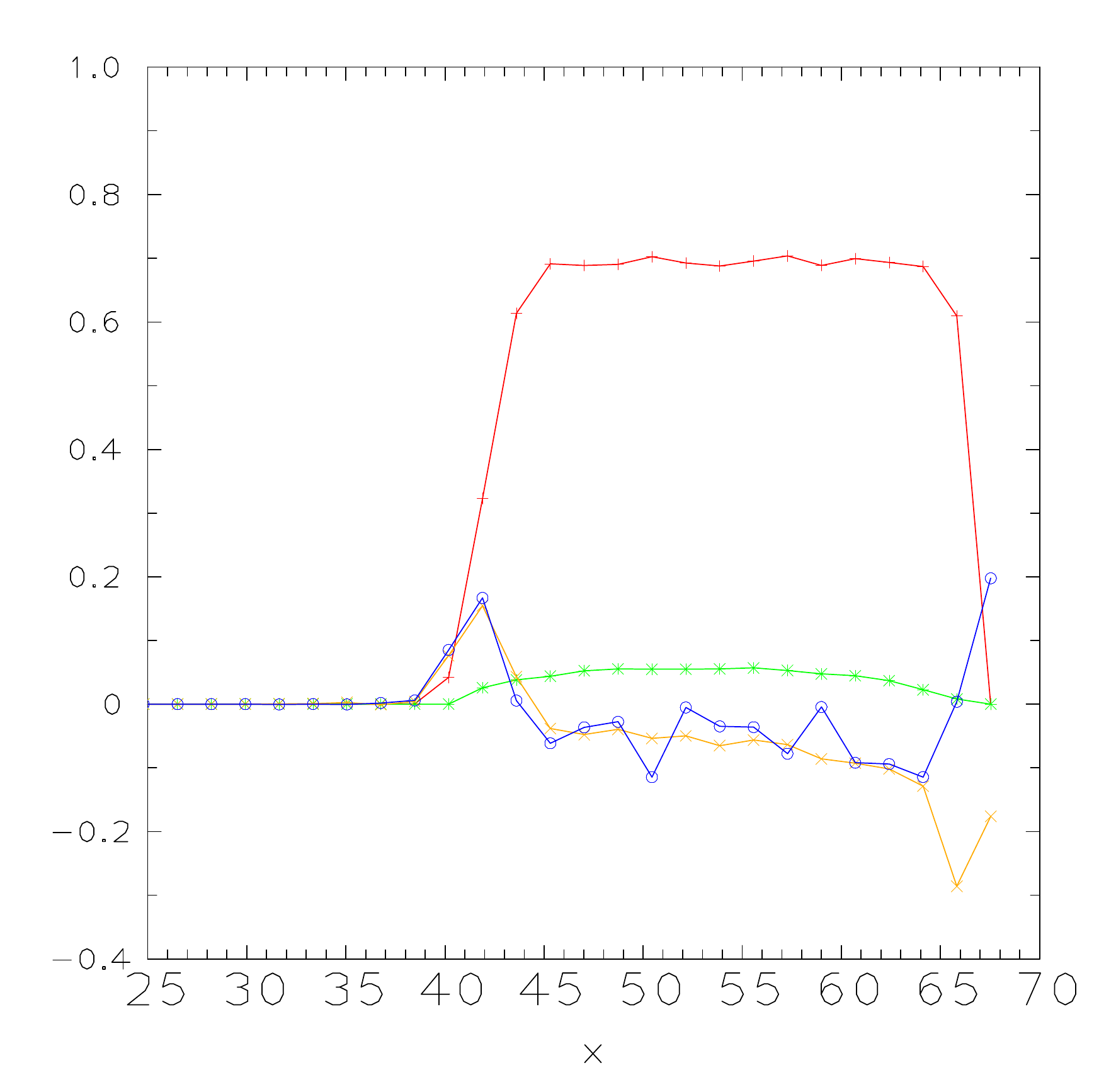}
\caption{\small{ Streamlines (top) and local fields (bottom) for a drop 
squeezed into a corner at $u=0.1$:  
density: red (+), $x$-velocity: green (*), $x$-force: orange (x) 
and $y$-force: blue ($\circ$). }}
\label{fxy-corner}
\end{figure}

We consider a range of wall velocities 0.01 to 1.0$\sigma/\tau$. 
At the lower velocities we use the same time step and wall tethering strength 
as in the previous simulations, but for the higher velocities the time step
is lowered to 0.001$\tau$ and instead of tethering thermally active wall
atoms, the wall is translated as a rigid (but slightly randomized) lattice.
The reason is that at higher velocities the atom positions lag significantly 
behind the tether positions unless a very high binding force is used, which
would require a correspondingly very small time step to resolve. We have
verified that the fluid motion and wall forces are not sensitive to this
modification. When the lower wall moves (right), the drop is squeezed into
the corner, as shown in Fig.~\ref{snap-corner} for the $u_w=0.1$ case, and
evolves to a roughly time-independent configuration after several hundred
$\tau$.  In the steady state, 
the advancing contact line at the horizontal wall (in the rest frame of
this wall the contact line advances to the left) increases to a
velocity-dependent value (100 to 145$^\circ$), while the static angle on 
the vertical wall shows little change.  The motion inside the drop is
rolling, as seen in Fig.~\ref{fxy-corner}, accompanied by some slip at the
wall. The horizontal force the liquid exerts on the on the wall is negative,
opposing the wall motion, with a peak at the corner where a continuum
no-slip flow field would have a stress singularity, while the vertical force
has the same interpretation as for sessile drops.
At higher wall velocities, the liquid density profile becomes
slightly asymmetric and larger near the corner, the slip velocity increases,
also asymmetrically but instead largest at the contact line, and the force 
profiles increase in magnitude but maintain the same shape.

\begin{figure}
\centering
\includegraphics[scale=0.40]{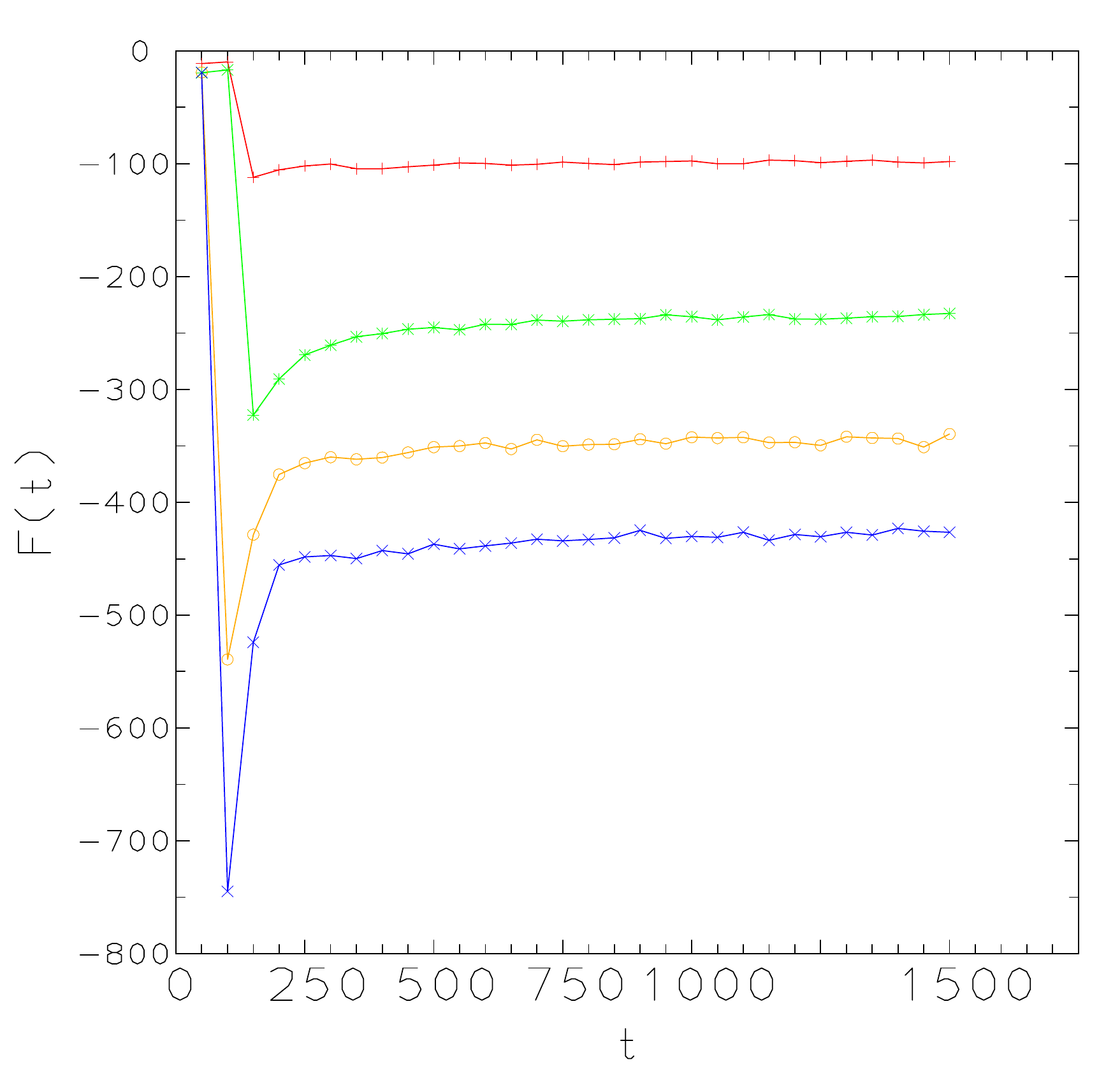}\hspace{0.1in}
\includegraphics[scale=0.40]{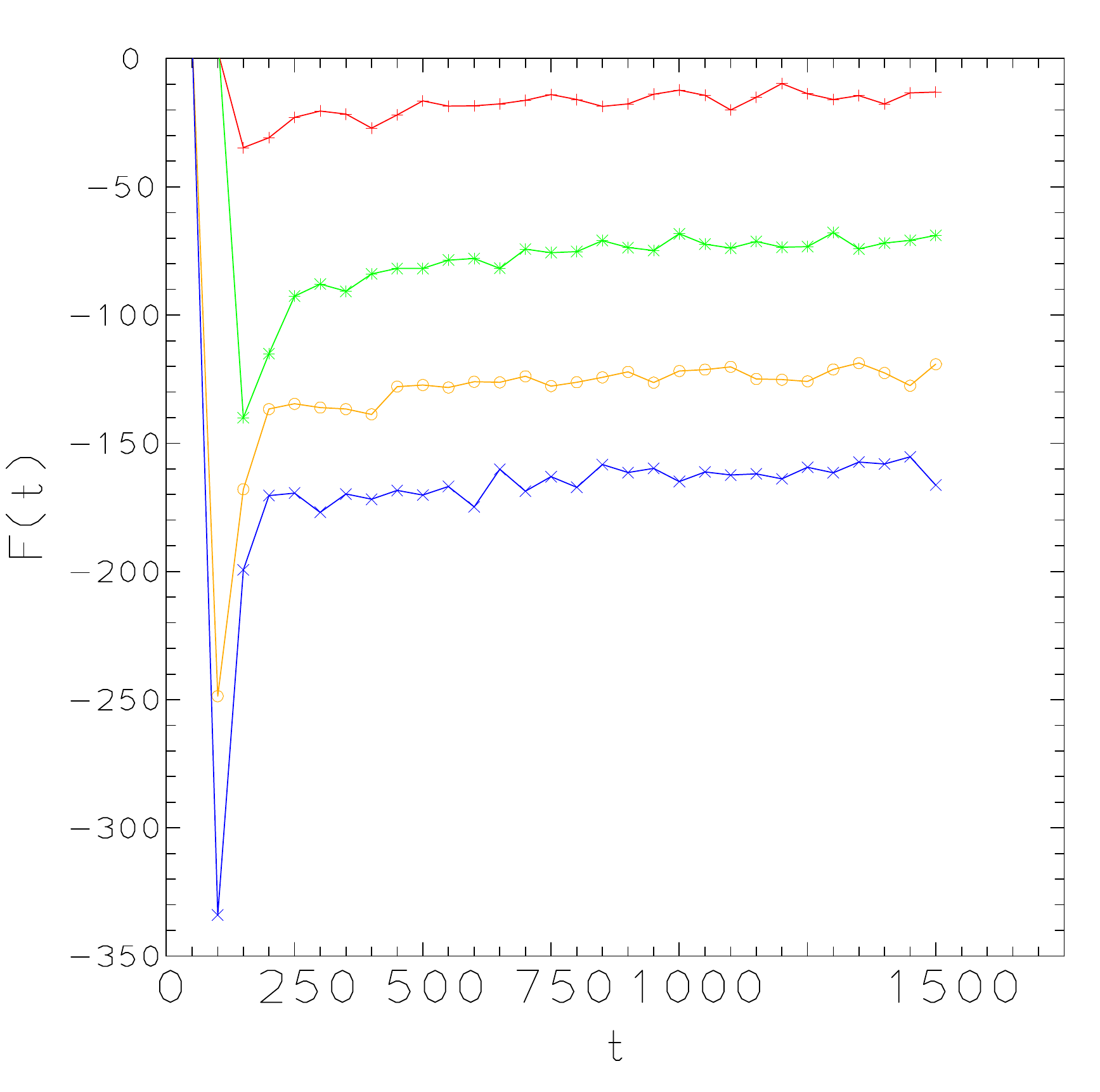}
\caption{\small{ Time variation of the $x$ (top) and $y$ (bottom) force 
on the sliding wall for various velocities. $u=0.1$: red (+); $u=0.25$:
green (*); $u=0.5$: orange ($\circ$) and $u=1.0$: blue (x). }} 
\label{cor-fxy}
\end{figure}

The $x$ and $y$ components of the total force on the bottom wall are shown
in Fig.~\ref{fxy-corner}, for different values of the wall velocity. In all
cases there is an initial peak, resembling that appearing in Couette flow,
which we attribute to the inertial effects of an impulsive start. The
reason is that the duration of the stress peaks is 100-150$\tau$ (note that
wall motion begins at 100$\tau$) which is comparable to the vorticity
diffusion time across the drop, $R^2/\mu$ with drop size $R\sim 35\sigma$
and viscosity $\mu=0.518m/(\sigma\tau$, which characterizes flow development. 
The signs of the fluid forces on the wall are negative because the walls are
squeezing the fluid to the right and upwards and the fluid resists this.  

\begin{figure}
\centering
\includegraphics[scale=0.40]{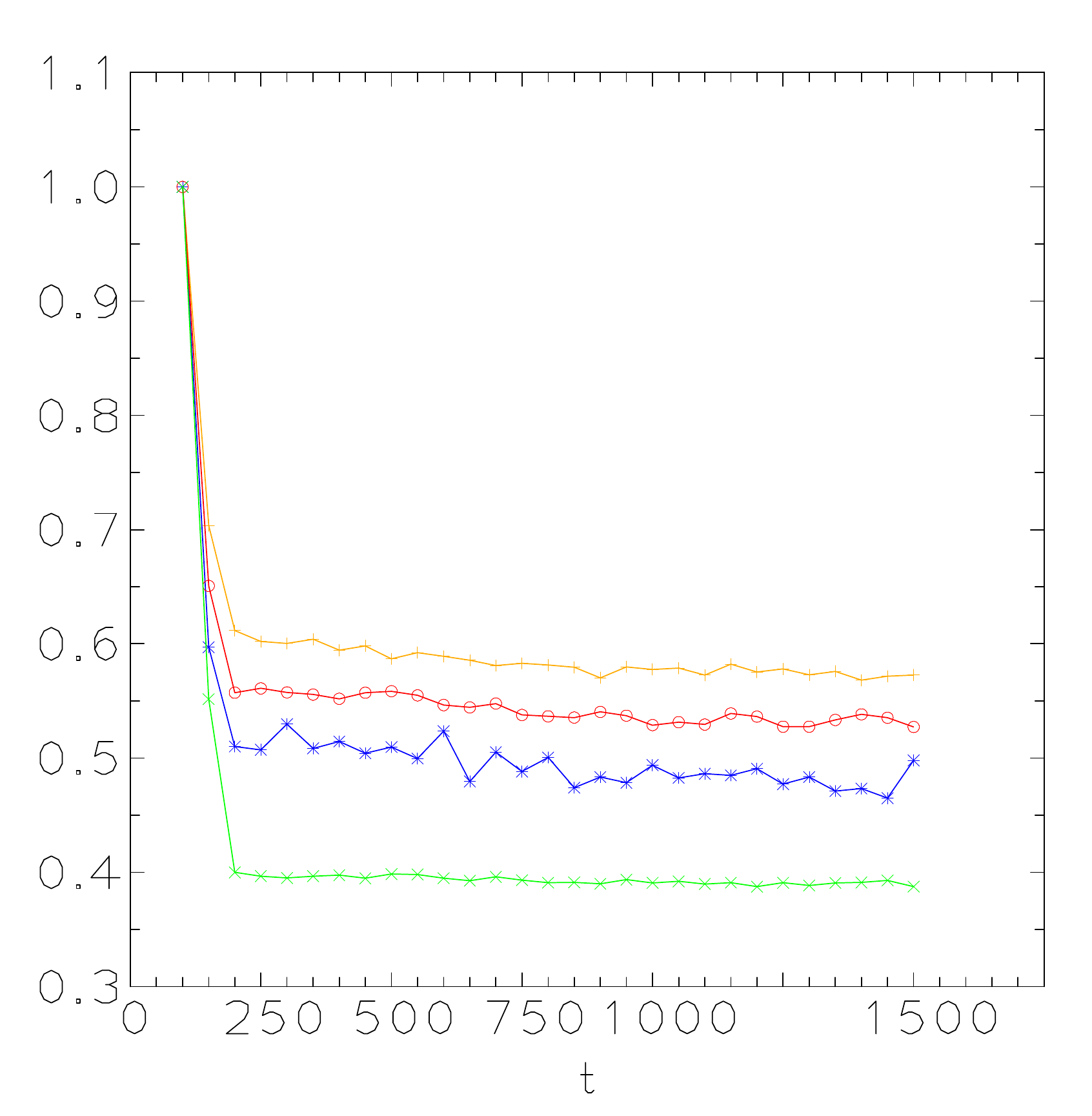}
\caption{\small{ Time decay of fluid density and velocity near the wall,
and the $x,y$ forces on the drop at $u=1.0$; color code as in Fig.~\ref{bottom}. }}  
\label{decay}
\end{figure}

The other notable feature of the forces is the
systematic weak decay (towards zero) following the peak, surprising since one 
would have 
expected constant wall forces in an apparently steady flow. However, in 
this situation the drop is not confined and is free to adjust its position.
intuitively one might imagine the drop pushed upwards by repeated collisions
with the sliding wall atoms and in fact precisely this behavior is observed.  
In Fig.~\ref{decay} we plot the density and velocity in the fluid adjacent
to the wall at low and high velocities, normalized to their values at time
when the wall begins to move.  The initial peak results from the abrupt
start, following which there is a weak decay of the density, meaning
fluid moves away from the wall, which produces parallel decays in the
fluid velocity and wall forces.  

\section{The Experiment}

There remains the question of interpreting the experiments of Gao et al.
The principal distinction between that experiment and the drop simulations
here is the distortion of the drop surface as it moves relative to the 
measuring pin. Rearrangement of the liquid drop's shape and the contact 
lines on the pin and on the substrate 
would certainly affect the force exerted  We have carried out several 
simulations on this process, involving a spherical cap drop and a finite-sized 
obstacle pin, rather than a barrier wall as in the previous section, but these 
simulations do not completely reproduce the experiments in terms of drop
distortion and wall force.  

\begin{figure}
\centering
\includegraphics[scale=0.20]{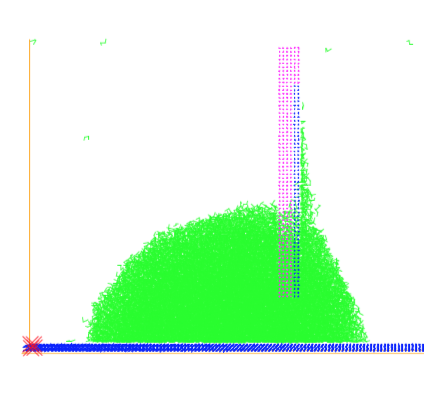}\hspace{0.1in}
\includegraphics[scale=0.20]{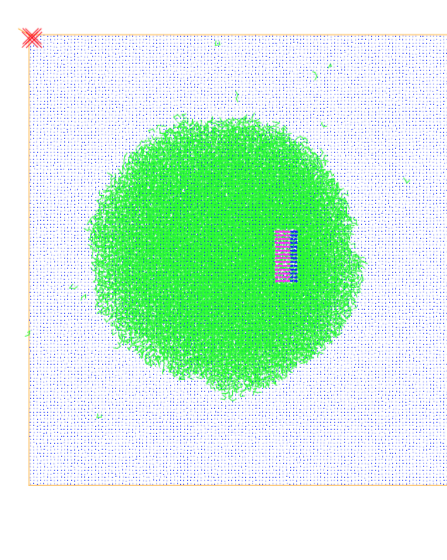}
\includegraphics[scale=0.20]{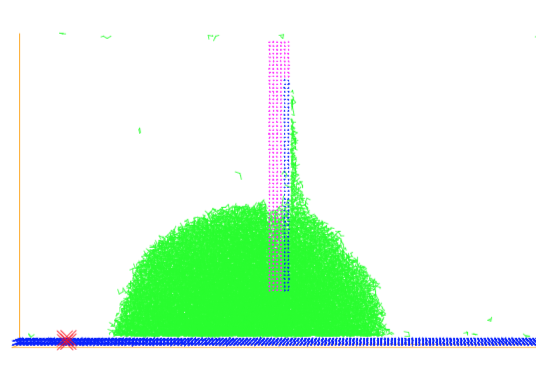}\hspace{0.1in}
\includegraphics[scale=0.20]{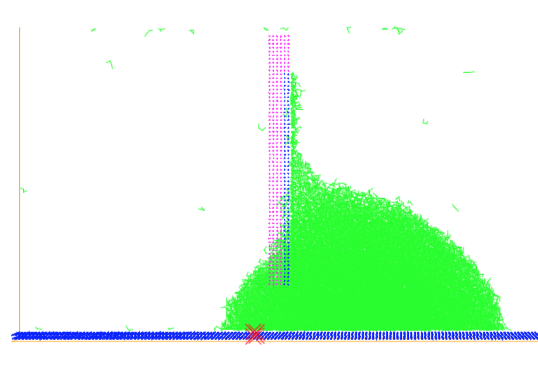}
\caption{\small{ Simulation of the experiment of Gao et al. \cite{gao}.
Top: side and top view of the equilibrated system, Bottom: effects of sliding
the substrate to the left, at times 1000 and 5000$\tau$. }}
\label{pin}
\end{figure}

\begin{figure}
\centering
\includegraphics[scale=0.40]{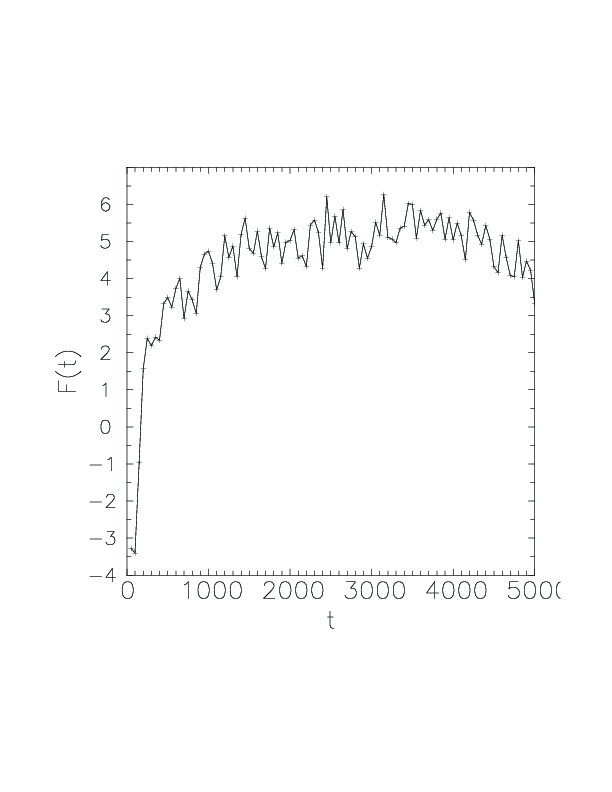}
\caption{\small{ Force on the pin in the simulation shown in
Fig.~\ref{pin}.}}
\label{pin-force}
\end{figure}

An example is given in Fig.~\ref{pin} -- a spherical cap is placed on a
partially wetting substrate ($c_{ff}=0.85$) and equilibrated with a 
rectangular pin inserted in the drop from above.  Most of the pin has the   
same interaction as the substrate, except that the upstream face is more
strongly wetting ($c_{ff}=1.2$), in order to mimic the experiment where a 
metallic reflecting layer was added to improve the imaging.
After equilibration the substrate is translated to the right at velocity
$u=0.1$ and the force on the pin recorded.  
Fig.~\ref{pin-force} again shows a monotonic rise to a near-plateau, with no 
``static-friction'' enhancement.  At times beyond 5000$\tau$ the
drop becomes highly elongated and eventually detaches from the pin, but
during the interval indicated the footprint of the drop on the substrate remains
approximately circular.  The results are similar when the conditions of the
simulation are varied (different wettabilities, different pin shapes,
different speeds, etc.), and {\em provided} the pin remains embedded inside
the drop the force on the pin is roughly constant.  Exceptions to the
typical behavior are found when the pin is at the edge of the drop, when
the liquid either first wets or dewets the pin. In these cases the
liquid/vapor interface does deform and a transient peak or spike appears in 
the force.

In contrast to the simulations, the experimental drops do change shape and, in
particular, their contact line length increases during the time interval 
when the force is enhanced. A further distinction in the simulations, a 
consequence of size limitations, is that the liquid climbing the wetting side 
of the pin has a density corresponding to liquid/vapor interfacial region
and does not faithfully represent bulk liquid and may not exert the proper
hydrodynamic drag.  Nonetheless, it appears that the transient friction 
enhancement in the experiments is related to change of drop shape, and is 
not a general characteristic of solid/liquid friction.

\section{Conclusions}

We have used MD simulations to investigate the possibility of an enhanced
shear stress before a liquid begins to move across a solid surface, analogous
to the distinction between static and dynamic friction when two solids move
relative to each other. In cases where the motion is driven by a constant 
body force applied to the interior of the liquid, even in a step-wise fashion, 
the stress is found to increase monotonically from zero to a steady state
value.  An abrupt motion of a solid bounding surface, however, generates a
large local strain which in turn produces a peak in the shear stress. Such
peaks are infinite in the (mathematical) continuum limit but regularized to 
large but finite values in MD simulations and in real life, but otherwise
entirely in accord with the Navier-Stokes equations. Furthermore, even in
situations where the liquid is pinned by inhomogeneities and requires a
minimum threshold force for continuous motion, no force enhancement is
found.  The experiments \cite{gao} which motivated this work appear to 
incorporate changes in drop shape and thereby involve interfacial dynamics
as well as wetting considerations, and do not provide evidence for an analog
of static solid/solid friction.

\end{document}